\documentclass[12pt]{article}
\pdfoutput=1
\usepackage[latin1]{inputenc}
\usepackage{amssymb}
\usepackage{amsmath}
\usepackage{amsthm}
\usepackage{amsfonts}
\usepackage[pdftex]{graphicx}
\usepackage{geometry}
\usepackage{braket}
\usepackage{enumerate}
\usepackage{subcaption}
\allowdisplaybreaks
\usepackage[table,usenames,dvipsnames]{xcolor}
\usepackage{setspace}
\usepackage[in]{fullpage}
\usepackage{hyperref}
\usepackage{array}
\usepackage{cancel}
\usepackage{wrapfig}
\usepackage[font=small,labelfont=bf]{caption}
\usepackage{mathtools}
\usepackage[nottoc,notlot,notlof]{tocbibind}
\usepackage[nosort]{cite}
\usepackage{parskip}

\usepackage{tikz}

\usetikzlibrary{calc,arrows, decorations.markings, shapes.misc, decorations.pathmorphing}

\def\boxsize{0.7}
\def\blobsize{0.55cm}
\def\bwblobsize{0.35cm}

\tikzset{
  ->-/.style={
    decoration={
      markings,
      mark=at position #1 with {\arrow{latex}}},
    postaction={decorate}
  },
  ->-/.default=0.5
}

\tikzset{
    wavy/.style={decorate, decoration={snake}, draw=red},
}

\tikzset{VO/.style={cross out, draw, 
         minimum size=5pt, 
         inner sep=0pt, outer sep=0pt}}

\tikzset{VOline/.style={decorate, decoration={snake}},
}

\tikzset{
    partial ellipse/.style args={#1:#2:#3}{
        insert path={+ (#1:#3) arc (#1:#2:#3)}
    }
}

\newcommand{\blackdot}{\node[circle, fill=black, draw, inner sep=0pt, minimum size=\bwblobsize]}
\newcommand{\whitedot}{\node[circle, fill=white, draw, inner sep=0pt, minimum size=\bwblobsize]}
\newcommand{\greyblob}{\node[circle, fill=black!20, draw, inner sep=2pt, minimum size=\blobsize]}

\newcommand{\drawULblack}{\blackdot (UL) at (-\boxsize,  \boxsize) {};}
\newcommand{\drawURblack}{\blackdot (UR) at ( \boxsize,  \boxsize) {};}

\newcommand{\drawLLblack}{\blackdot (LL) at (-\boxsize, -\boxsize) {};}

\newcommand{\drawULwhite}{\whitedot (UL) at (-\boxsize,  \boxsize) {};}
\newcommand{\drawURwhite}{\whitedot (UR) at ( \boxsize,  \boxsize) {};}

\newcommand{\drawLLwhite}{\whitedot (LL) at (-\boxsize, -\boxsize) {};}

\newcommand{\drawUL}[1]{\greyblob (UL) at (-\boxsize,  \boxsize) {#1};}
\newcommand{\drawUR}[1]{\greyblob (UR) at ( \boxsize,  \boxsize) {#1};}
\newcommand{\drawLR}[1]{\greyblob (LR) at ( \boxsize, -\boxsize) {#1};}

\newcommand{\drawULempty}{\coordinate (UL) at (-\boxsize,  \boxsize);}
\newcommand{\drawURempty}{\coordinate (UR) at ( \boxsize,  \boxsize);}
\newcommand{\drawLRempty}{\coordinate (LR) at ( \boxsize, -\boxsize);}
\newcommand{\drawLLempty}{\coordinate (LL) at (-\boxsize, -\boxsize);}

\newcommand{\drawboxinternallines}{
  \draw (UL) -- (UR) -- (LR) -- (LL) -- (UL);
}

\newcommand{\drawregionvariables}[5]{
  \draw ( 0.0, -1.4) node {#1};
  \draw (-1.4,  0.0) node {#2};
  \draw ( 0.0,  1.4) node {#3};
  \draw ( 1.4,  0.0) node {#4};
  \draw ( 0.0,  0.0) node {#5};
}

\newcommand{\drawuppercut}{
\draw (blobL) to[out=60, in=120] node[circle, fill=white, draw=white]{} (blobR);
\draw[red, dashed] (0, 0.5) -- (0, 0.9);
}

\newcommand{\drawlowercut}{
\draw (blobL) to[out=-60, in=-120] node[circle, fill=white, draw=white]{} (blobR);
\draw[red, dashed] (0,-0.5) -- (0,-0.9);
}

\numberwithin{equation}{section}

\hypersetup{
	colorlinks=true,
	linktoc=page,
	citecolor=Blue,
	linkcolor=Blue,
	urlcolor=Blue} 

\makeatletter
\renewcommand{\maketitle}
{ \begingroup \begin{center} \large {\bf \@title}
		\vskip 5pt \large \@author \\ \vskip 5pt \@date \end{center}
	\vskip 5pt \endgroup \setcounter{footnote}{0} }
\makeatother

\def\d{\delta}\def\D{\Delta}
\def\eps{\epsilon}
\DeclareMathOperator{\Li}{Li_2}

\long\def\symbolfootnote[#1]#2{\begingroup%
	\def\thefootnote{\fnsymbol{footnote}}\footnote[#1]{#2}\endgroup}

\begin{document}
	
	\begin{flushright}
		QMUL-PH-18-33\\
	\end{flushright}
	
	\vspace{20pt}
	
	\begin{center}
		
		{\Large \bf    Dual conformal invariance for form factors}
		
		\vspace{25pt}
		
		{\mbox {\bf  Lorenzo~Bianchi, Andreas~Brandhuber, 
				Rodolfo Panerai and  %
				Gabriele~Travaglini$^{\S}$}}%

		\vspace{0.5cm}
		
		\begin{center}
			{\small \em
						Centre for Research in String Theory\\
						School of Physics and Astronomy\\
						Queen Mary University of London\\
						Mile End Road, London E1 4NS, United Kingdom
			}
		\end{center}
		
		\vspace{15pt}
		
{\bf Abstract}
	\end{center}
	\vspace{0.3cm} 
	
\noindent
	
\noindent
Form factors of the stress-tensor multiplet operator in $\mathcal{N}\!=\!4$ supersymmetric Yang-Mills reveal surprisingly simple structures similar to those appearing in  scattering amplitudes. In this paper we show that, as for the case of amplitudes, they also enjoy  dual conformal symmetry. We compute the dual conformal anomaly at one loop for an arbitrary number of particles and generic helicities, which matches the expression of the dual conformal anomaly of amplitudes, and perform explicit checks for MHV and NMHV one-loop form factors. In the NMHV case the realisation of dual conformal symmetry  requires a delicate cancellation of offending terms arising from three-mass triangles, which we explicitly check in the case of the four-point NMHV form factor.

	\vfill
	\hrulefill
	\newline
\vspace{-1cm}
$^{\S}$~\!\!{\tt\footnotesize\{lorenzo.bianchi, a.brandhuber, r.panerai, g.travaglini\}@qmul.ac.uk}	
	
	\setcounter{page}{0}
	\thispagestyle{empty}
	\newpage
	
	\setcounter{tocdepth}{4}
	\hrule height 0.75pt
	\tableofcontents
	\vspace{0.8cm}
	\hrule height 0.75pt
	\vspace{1cm}
	
	\setcounter{tocdepth}{2}	

	\newpage

\section{Introduction}
Dual conformal symmetry is a highly non-trivial feature of scattering amplitudes in $\mathcal{N}\!=\!4$ super Yang-Mills (SYM) theory. Historically, it was first noticed that the integrals appearing in the perturbative expansion of the four-point amplitude enjoy conformal invariance when expressed in terms of dual variables \cite{Drummond:2006rz,Drummond:2007aua}. More precisely, they would be dual conformal invariant if they could be computed in four dimensions. The need for an  infrared (IR) regulator breaks dual conformal invariance and generates an anomaly \cite{Drummond:2007au,Brandhuber:2008pf}, which  is however  under complete control \cite{Drummond:2007au} and at one loop induces relations among the supercoefficients of the box integrals entering the final result \cite{Brandhuber:2009xz,Korchemsky:2009hm}. Moreover, a one-loop unitarity-based derivation of this anomaly  for arbitrary helicities  and number of external legs was presented  in \cite{Brandhuber:2009kh}. 

It soon  became also  clear that tree-level scattering amplitudes are invariant under the full dual superconformal group \cite{Drummond:2008vq} and the symmetry can be extended to an infinite dimensional Yangian algebra \cite{Drummond:2009fd}. Since even at tree level the full amplitude is, strictly speaking, only covariant, not invariant, under dual conformal transformations, it is convenient to work with ratios of amplitudes. In practice, one usually divides the result by the tree-level MHV amplitude -- something that we find more natural to do also in our work -- and the resulting  ratio is then  invariant up to anomalies due to IR divergences. A convenient way to show this invariance is to introduce momentum twistors \cite{Mason:2009qx}. These variables allow dual superconformal transformations to act linearly, and are helpful to systematically construct superconformal invariants \cite{Korchemsky:2010ut}.
More recently, dual conformal symmetry received renewed attention. On the one hand, the authors of \cite{Bourjaily:2013mma} developed an IR regulator making dual conformal invariance of finite observables manifest at the integrand level, on the other hand a careful analysis has shown the emergence of hidden symmetries in the non-planar sector of amplitudes  \cite{Ben-Israel:2018ckc,Bern:2018oao,Chicherin:2018wes,Bern:2018ido}.

In this paper, we want to extend the notion of dual conformal symmetry to form factors of the stress tensor multiplet operator in  $\mathcal{N}\!=\!4$ SYM theory. Form factors of half-BPS operator  are by now very well studied,   both at weak \cite{vanNeerven:1985ja,Brandhuber:2010ad,Brandhuber:2011tv,Gehrmann:2011xn,Bork:2011cj,Brandhuber:2012vm,Bork:2012tt,Boels:2015yna,Boels:2017ftb,Meidinger:2017hvm} and strong coupling \cite{Alday:2007he,Maldacena:2010kp}. The extension to form factors of the on-shell diagram formalism and their formulation in terms of twistor variables, exhibiting an underlying Gra{\ss}mannian geometry, have also been studied  \cite{Bork:2014eqa,Frassek:2015rka,Koster:2016fna,Koster:2016loo,Koster:2016ebi,Bork:2016hst,Brandhuber:2016xue}. Yet, despite the availability of  many perturbative results, the dual conformal symmetry properties of  form factors of protected operators have not yet been investigated (see \cite{Bork:2015fla} for comments regarding the $q^2=0$ case). 
One reason why this question was not addressed is the presence of triangle integrals in the expressions for one-loop form factors. 

Triangles, unlike boxes,  are expected to break dual conformal invariance  explicitly, as one can see  easily. Consider first a one-loop box integral in dual variables, which is  given by
\begin{equation}
I_4= \int \mathrm{d}^4 x_0 \; \frac{1}{x_{01}^2 x_{02}^2 x_{03}^2 x_{04}^2}\ . 
\end{equation}
Performing an inversion $x_i\to {x_i}/{x_i^2}$ and a compensating change of variables $x_0\to {x_0}/{x_0^2}$ (which implies $\mathrm{d}^4 x_0  \to \mathrm{d}^4 x_0/x_0^8$) one gets
\begin{equation}
 I_4 \to I_4 \, x_1^2 \, x_2^2 \, x_3^2 \, x_4^2 \ , 
\end{equation}
which can be compensated by a numerator $x_{13}^2 x_{24}^2$. This is not possible for the triangle 
integral
\begin{equation}
 I_3= \int \mathrm{d}^4 x_0 \; \frac{1}{x_{01}^2 x_{02}^2 x_{03}^2} \ , 
\end{equation}
whose integrand variation  depends explicitly on the loop variable $x_0$, preventing a covariant transformation. This led to the expectation that any quantity involving triangle integrals cannot be dual conformal invariant. We will show in the following that this expectation is naive, and our careful analysis of the form factors at tree (one-loop) level will reveal the presence of (anomalous) dual conformal symmetry in complete analogy to the case of amplitudes.
We will start from  tree level,  where dual conformal invariance descends from the invariance of certain $R$-functions appearing in tree-level form factors. 
We then move to one loop, where we present a derivation of the dual conformal anomaly along the lines of  \cite{Brandhuber:2009kh}, 
and importantly also explicitly check the dual conformal anomaly for the MHV and NMHV~cases. 

A key aspect of our investigation is the appropriate assignment of region variables for form factors introduced in \cite{Part1}. For the case of scattering amplitudes, the sum of external on-shell momenta vanishes and dual momenta are the vertices of a light-like polygon. For form factors, the presence of the operator insertion leads one to consider  a periodic configuration of region variables \cite{Alday:2007he, Brandhuber:2010ad}. In the following, we will describe an unambiguous prescription to assign region variables for a given ordering of the external legs. Note however that special conformal transformations do not preserve distances, and consequently do not preserve periodicity under translations. In general, a periodic configuration of the dual variables is invariant under a discrete translation by a period $q$. We denote by $\mathbb{P}$ the action of such a  translation. After a dual special conformal transformation $\mathcal{K}$, the configuration will be invariant under the action of twisted periodicity
\begin{align}
\tilde{\mathbb{P}} = \mathcal{K} \cdot \mathbb{P} \cdot \mathcal{K}^{-1}  \;.
\end{align}
This subtlety  was already noticed  in \cite{Ben-Israel:2018ckc}, where the authors looked at double-trace scattering amplitudes and argued that the original Wilson line correlator and the twisted one correspond to the same scattering amplitude. Here we find a very similar picture: a dual conformal transformation maps a configuration of region variables, which is periodic under translations,  to a configuration that obeys twisted periodicity; nevertheless, we will show that this does not change the final result of the tree-level form factor (or to be more precise the appropriate ratio), and at one loop  induces an anomaly
that is completely analogous to that  of amplitudes.


The rest of the paper is organised as follows. In Section \ref{tree_level}, we review the tree-level results of \cite{Bork:2014eqa}, with a particular focus on dual conformal symmetry, made manifest by the formulation in terms of twistor variables. In Section \ref{oneloop}, we provide a unitarity-based derivation  of the anomalous  dual conformal symmetry at one loop. We then test our findings in Section \ref{examples}, where we show explicitly that MHV and NMHV one-loop form factors obey the same anomalous dual-conformal Ward identity as amplitudes. Several technical details and definitions are included in four appendices.

\section{Dual conformal symmetry of  tree-level form factors}\label{tree_level}

As for the case of scattering amplitudes, it is convenient to analyse the properties of the ratio $\tilde{F}^{(0)}_{n,k}$ defined as
\begin{align}\label{ratio}
 \tilde{F}^{(0)}_{n,k}=\frac{F^{(0)}_{n,k}}{F^{(0)}_{n,0}}\,,
\end{align}
where, in our notation, $F^{(l)}_{n,k}$ is the $n$-point N$^k$MHV form factor at $l$ loops (see Appendix~\ref{conventions} for our conventions). We will show that the ratio $\tilde{F}^{(0)}_{n,k}$ is invariant under dual conformal transformations. This feature was already mentioned in \cite{Bork:2014eqa}, and  here we review some of the results of that paper, focusing on the properties under dual conformal transformations. 

We start by reviewing some facts about scattering amplitudes. It was noticed in \cite{Drummond:2008vq,Brandhuber:2008pf} that the ratio $A^{0}_{n,1}/A^{(0)}_{n,0}$ can be expressed as a linear combination of dual conformal invariant objects, called $R$-invariants. It  was then realised, using supersymmetric recursion relations \cite{ArkaniHamed:2008gz, Brandhuber:2008pf}, that all tree-level amplitudes in $\mathcal{N}\!=\!4$ SYM can be expressed as combinations of $R$-invariants \cite{Drummond:2008vq,Drummond:2008cr}. The latter can be related to four-particle cuts of one-loop amplitudes, thus establishing important relations between loops and trees \cite{Britto:2004nc,Britto:2005fq,Drummond:2008bq}. The $R$-invariants can be defined for an arbitrary assignment of external region variables as
\begin{align}\label{Rfunction}
\begin{tikzpicture}[thick, scale=0.8]
 \node[left] at (-2.5,0) {$R_{rst}=$};
  \drawLLwhite
  \drawUL{$\scriptstyle{0}$}
  \drawUR{$\scriptstyle{0}$}
  \drawLR{$\scriptstyle{0}$}
   \drawregionvariables{$x_c$}{$x_{c+1}$}{$x_a$}{$x_b$}{}
  \drawboxinternallines
  \draw (UL) -- ++( 180:0.8) node[anchor=east] {$r+1$};
  \draw[dotted] (UL)+( 170:0.6) to [bend left=45] ++( 100:0.6);
  \draw (UL) -- ++(  90:0.8) node[anchor=south] {$s-1$};
  \draw (UR) -- ++( 0:0.8) node[anchor= west] {$t-1$};
  \draw[dotted] (UR)+(  90:0.6) to [bend left=45] ++(  0:0.6);
  \draw (UR) -- ++(  90:0.8) node[anchor=south] {$s$};
  \draw (LR) -- ++(   0:0.8) node[anchor=west] {$t$};
  \draw[dotted] (LR)+( -10:0.6) to [bend left=45] ++( -80:0.6);
  \draw (LR) -- ++( -90:0.8) node[anchor=north] {$r-1$}; 
  \draw (LL) -- ++(-135:0.8) node[anchor=north east] {$r$};
  \node[right] at (2.5,0) {$\displaystyle=\frac{\langle s-1\,s\rangle \langle t-1\,t\rangle \, \delta^{(4)}(\langle r|x_{ca}x_{ab}|\theta_{bc}\rangle  + \langle r|x_{cb}x_{ba}|\theta_{ac}\rangle )}{x_{ab}^2 \langle r|x_{cb}x_{ba}|s-1\rangle  \langle r|x_{cb}x_{ba}|s\rangle  \langle r|x_{ca}x_{ab}|t-1\rangle  \langle r|x_{ca}x_{ab}|t\rangle }$}; 
\, .
\end{tikzpicture}
\end{align}
We begin by showing that this function is invariant under dual conformal transformations and then we will discuss how to adapt this construction to the case of form factors. 
We also introduce the four-bracket
\begin{align}\label{fourbrack}
 \braket{i,j-1,j,k}=\langle i|x_{ij}x_{jk}|k\rangle \, \langle j-1 \, j\rangle \ , 
\end{align}
 and notice that, since $p_i=x_i-x_{i+1}$,  the following identity
\begin{align}
 x_{i}\ket{i}=x_{i+1} \ket{i} 
\end{align}
holds. 
Therefore, we can replace  $x_i$ in \eqref{fourbrack} by $x_{i+1}$, and $x_k$ by $x_{k+1}$. The crucial requirement is that $x_i$ and $x_k$ label one of the two regions adjacent to $p_i$ and $p_k$, respectively. 

The easiest way to see that the combination \eqref{fourbrack} is invariant is by introducing momentum twistors \cite{Mason:2009qx}
\begin{align}
 Z^{\hat{A}}_i&=\begin{pmatrix} \lambda_i^\alpha\\ \mu_i^{\dot \alpha}\end{pmatrix} \, , & \mu_i^{\dot \alpha}=x_i^{\dot \alpha \alpha} \lambda_{i \alpha}\, .
\end{align}
In these variables, conformal transformations act linearly. In particular, they are implemented as an $\mathrm{SL}(4)$ transformation on the index $\hat{A}$. The four-bracket \eqref{fourbrack} is defined as
\begin{equation}
  \braket{i,j-1,j,k}= \epsilon_{\hat A \hat B \hat C \hat D} Z^{\hat A}_i Z^{\hat B}_{j-1} Z^{\hat C}_j Z^{\hat D}_k\, , 
\end{equation}
and it is therefore manifestly invariant under $\mathrm{SL}(4)$. It is also convenient to introduce  supertwistor variables 
\begin{align}
 \mathcal{Z}_i^M&=\begin{pmatrix} Z_i^{\hat A} \\ \chi_i^{A}\end{pmatrix} \ ,&  \chi^A_i=\theta_i^{A\, \alpha} \lambda_{i \alpha}\ , 
\end{align}
transforming in the fundamental representation of the supergroup $\mathrm{SL}(4|4)$. Its projective real section $\mathrm{PSU}(2,2|4)$ is precisely the $\mathcal{N}\!=\!4$ superconformal group.
It is easy to see that, given five arbitrary superstwistors $\mathcal{Z}_a,\ldots, \mathcal{Z}_e$, the quantity
\begin{align}\label{fivebrack}
[a,b,c,d,e]= \frac{\d^{(4)}(\braket{a,b,c,d} \chi_{e}+\text{cyclic})}{\braket{a,b,c,d}\braket{b,c,d,e}\braket{c,d,e,a}\braket{d,e,a,b}\braket{e,a,b,c}}
\end{align}
is an $\mathrm{SL}(4|4)$ invariant. Furthermore, \eqref{fivebrack} is invariant under an arbitrary rescaling
\begin{align}\label{scaling}
\mathcal{Z}^{M}_i\to \zeta_i \mathcal{Z}^{M}_i  \ , 
\end{align}
which is related to the little group scaling. 
This is a condition  that must  be satisfied given the projective nature of twistor variables.
After some manipulations one can show that the $R$-invariant is just a specific instance of this general invariant \cite{Mason:2009qx}:
\begin{align}
 R_{rst}=[s-1,s,t-1,t,r] \ .
\end{align}
An important difference between the amplitude and the form factor computation is that there is no momentum conservation for the external legs, {\it i.e.} 
\begin{align}
 \sum_{i=1}^{n} p^{\alpha \dot \alpha}_i&=q^{\alpha \dot \alpha}\ , & \sum_{i=1}^{n} \mathrm{q}^{A\alpha}_i &=\gamma^{A\alpha} \ , 
\end{align}
and 
\begin{align}
 p^{\alpha \dot \alpha}_i&=\lambda_i^{\alpha} \tilde\lambda_i^{\dot \alpha} \ ,  &  \mathrm{q}^{A\alpha}_i&=\eta_i^A \lambda_i^{\alpha}\ .
\end{align} 
Consequently, region supermomenta are defined on a periodic contour \cite{Alday:2007he,Brandhuber:2010ad,Brandhuber:2011tv}
\begin{align}
 x^{\alpha\dot\alpha}_i&\sim x^{\alpha\dot\alpha}_i  + m  q^{\alpha\dot\alpha}  \ , & \theta_i^{A\alpha}&\sim \theta_i^{A\alpha} + m \gamma^{A\alpha} \ ,
\end{align}
for $m\in \mathbb{Z}$.
This introduces a redundancy in the assignment of dual variables and one has to establish a consistent convention. This issue was already discussed in \cite{Bork:2014eqa,Part1}. Here, we follow the convention of \cite{Part1}, which can be summarised as follows. We choose one particular period, whose points are called $(x_i,\theta_i)$. Image points  belonging to the other periods  are indicated using the notation
\begin{align}
 x_i^{[m]}&=x_i+m\, q \ , & \theta_i^{[m]}=\theta_i+m\, \gamma\ .
\end{align}
For the specific case $m=\pm 1$, we also use $x_i^{\pm}=x_i\pm q$ and $\theta_i^{\pm}=\theta_i\pm  \gamma$. Notice that, for any $m\in \mathbb{Z}$,
\begin{align}\label{supermomenta}
 p_i&=x_i^{[m]}-x_{i+1}^{[m]} \ , & \mathrm{q}_i=\theta_i^{[m]}-\theta_{i+1}^{[m]}\ .
\end{align}

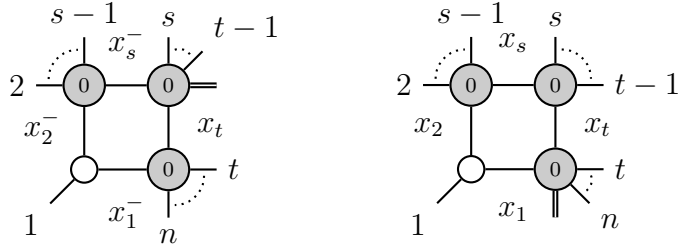
\begin{figure}[htbp]
\centering
\begin{tikzpicture}[thick, scale=0.8]
  \drawLLwhite
  \drawUL{$\scriptstyle{0}$}
  \drawUR{$\scriptstyle{0}$}
  \drawLR{$\scriptstyle{0}$}
  \drawboxinternallines
  \drawregionvariables{$x_1^-$}{$x_2^-$}{$x_s^-$}{$x_t$}{}
  \draw (UL) -- ++( 180:0.8) node[anchor=east] {$2$};
  \draw[dotted] (UL)+( 170:0.6) to [bend left=45] ++( 100:0.6);
  \draw (UL) -- ++(  90:0.8) node[anchor=south] {$s-1$};
  \draw[double] (UR) -- ++(   0:0.8) node[] {};
  \draw (UR) -- ++(  45:0.8) node[anchor=south west] {$t-1$};
  \draw[dotted] (UR)+(  80:0.6) to [bend left=45] ++(  50:0.6);
  \draw (UR) -- ++(  90:0.8) node[anchor=south] {$s$};
  \draw (LR) -- ++(   0:0.8) node[anchor=west] {$t$};
  \draw[dotted] (LR)+( -10:0.6) to [bend left=45] ++( -80:0.6);
  \draw (LR) -- ++( -90:0.8) node[anchor=north] {$n$}; 
  \draw (LL) -- ++(-135:0.8) node[anchor=north east] {$1$};
\end{tikzpicture}\hspace{1cm}
\raisebox{.03\height}{
\begin{tikzpicture}[thick, scale=0.8]
  \drawLLwhite
  \drawUL{$\scriptstyle{0}$}
  \drawUR{$\scriptstyle{0}$}
  \drawLR{$\scriptstyle{0}$}
   \drawregionvariables{$x_1$}{$x_2$}{$x_s$}{$x_t$}{}
  \drawboxinternallines
  \draw (UL) -- ++( 180:0.8) node[anchor=east] {$2$};
  \draw[dotted] (UL)+( 170:0.6) to [bend left=45] ++( 100:0.6);
  \draw (UL) -- ++(  90:0.8) node[anchor=south] {$s-1$};
  \draw (UR) -- ++(   0:0.8) node[anchor=west] {$t-1$};
  \draw[dotted] (UR)+(  80:0.6) to [bend left=45] ++(  10:0.6);
  \draw (UR) -- ++(  90:0.8) node[anchor=south] {$s$};
  \draw (LR) -- ++(   0:0.8) node[anchor=west] {$t$};
  \draw[dotted] (LR)+( -10:0.6) to [bend left=45] ++( -40:0.6);
  \draw (LR) -- ++( -45:0.8) node[anchor=north west] {$n$};
  \draw[double] (LR) -- ++( -90:0.8) node[] {};
  \draw (LL) -- ++(-135:0.8) node[anchor=north east] {$1$};
\end{tikzpicture}
}
\caption{\it Examples of region variables assignment for two  $R$-invariants. We label region momenta starting from the region adjacent to the corner containing the   off-shell leg in clockwise order. 
}
\label{exassign}
\end{figure}

In extending the computation of $R$-invariants to form factors the off-shell leg appears in one of the MHV blobs in \eqref{Rfunction}. As done in \cite{Part1}, we use the position of the off-shell leg to start assigning region momenta and we ask that the first region we encounter always sits in the particular period we selected ({\it i.e.}~that with regions $x_i$). In the case of \mbox{$R$-invariants}, it is easy to understand how this works looking at Figure \ref{exassign}, where we selected two specific $R$-invariants with $r\!=\!1$, and we assigned  region variables accordingly. In Section~\ref{oneloop} we will use the same prescription for the case of one-loop form factors.

It should be clear that this is just one specific choice, we may well choose any other period but the 
result for any $R$-invariant would be unchanged. 
We stress that, as  discussed in the Introduction, dual special conformal transformations act differently for different periods,  and this  causes ambiguities in the  action on an MHV prefactor -- which is why we prefer to divide it out and work with quantities written in the form 
of $R$-invariants (see Section \ref{oneloop} for a discussion of the loop level case), and translating them in twistor variables as was done in \cite{Bork:2014eqa}. Also twistor variables are arranged in periodic configurations
\begin{align}\label{periodictwistors}
 \mathcal{Z}^{[m] M}_i&=\begin{pmatrix} Z_i^{[m] \hat A} \\ \chi_i^{[m] A} \end{pmatrix}\, , & Z_i^{[m]\hat A}&=\begin{pmatrix} \lambda_i^{\alpha} \\ (x_i^{[m]})^{\dot \alpha \alpha} \lambda_{i\alpha} \end{pmatrix}\, ,  & \chi_i^{[m] A}=(\theta_i^{[m]})^{A\alpha} \lambda_{i \alpha}\, ,
\end{align}
but this does not affect the invariance of \eqref{fivebrack}, which holds for five arbitrary twistors. This implies that whenever a result can be written in terms of five-brackets \eqref{fivebrack}, it is automatically invariant. 
Notice also that under rescaling \eqref{scaling}, for any $m\in \mathbb{Z}$,
\begin{align}\label{scalingperiodic}
 \mathcal{Z}^{[m] M}_i\to \zeta_i \, \mathcal{Z}^{[m] M}_i\, .
\end{align}
This can be understood by thinking of the rescaling \eqref{scaling} as a  freedom in the definition of $\lambda_i$. Since $\lambda_i$ is not affected by the shifts \eqref{periodictwistors}, all the image twistors should be rescaled by the same factor.

As we mentioned, in the generalisation to form factors, one of the MHV amplitudes in \eqref{Rfunction} is replaced by an MHV form factor. In \cite{Bork:2014eqa} it was shown that two different configurations are needed to compute the NMHV form factor. They are represented by
\begin{align}\label{FFRfunctions}
R'_{rst} =
\raisebox{-.48\height}{
\begin{tikzpicture}[thick, scale=0.8]
  \drawLLwhite
  \drawUL{$\scriptstyle{0}$}
  \drawUR{$\scriptstyle{0}$}
  \drawLR{$\scriptstyle{0}$}
  \drawboxinternallines
  \draw (UL) -- ++( 180:0.8) node[anchor=east] {$r+1$};
  \draw[dotted] (UL)+( 170:0.6) to [bend left=45] ++( 100:0.6);
  \draw (UL) -- ++(  90:0.8) node[anchor=south] {$s-1$};
  \draw[double] (UR) -- ++(   0:0.8) node[] {};
  \draw (UR) -- ++(  45:0.8) node[anchor=south west] {$t-1$};
  \draw[dotted] (UR)+(  80:0.6) to [bend left=45] ++(  50:0.6);
  \draw (UR) -- ++(  90:0.8) node[anchor=south] {$s$};
  \draw (LR) -- ++(   0:0.8) node[anchor=west] {$t$};
  \draw[dotted] (LR)+( -10:0.6) to [bend left=45] ++( -80:0.6);
  \draw (LR) -- ++( -90:0.8) node[anchor=north] {$r-1$}; 
  \draw (LL) -- ++(-135:0.8) node[anchor=north east] {$r$};
\end{tikzpicture}
} \;, \qquad
R''_{rst} =
\raisebox{-.46\height}{
\begin{tikzpicture}[thick, scale=0.8]
  \drawLLwhite
  \drawUL{$\scriptstyle{0}$}
  \drawUR{$\scriptstyle{0}$}
  \drawLR{$\scriptstyle{0}$}
  \drawboxinternallines
  \draw (UL) -- ++( 180:0.8) node[anchor=east] {$r+1$};
  \draw[dotted] (UL)+( 170:0.6) to [bend left=45] ++( 100:0.6);
  \draw (UL) -- ++(  90:0.8) node[anchor=south] {$s-1$};
  \draw (UR) -- ++(   0:0.8) node[anchor=west] {$t-1$};
  \draw[dotted] (UR)+(  80:0.6) to [bend left=45] ++(  10:0.6);
  \draw (UR) -- ++(  90:0.8) node[anchor=south] {$s$};
  \draw (LR) -- ++(   0:0.8) node[anchor=west] {$t$};
  \draw[dotted] (LR)+( -10:0.6) to [bend left=45] ++( -40:0.6);
  \draw (LR) -- ++( -45:0.8) node[anchor=north west] {$r-1$};
  \draw[double] (LR) -- ++( -90:0.8) node[] {};
  \draw (LL) -- ++(-135:0.8) node[anchor=north east] {$r$};
\end{tikzpicture}
} \;,
\end{align}
and the expression of the $n$-point NMHV form factor is 
\begin{equation}\label{genNHMV}
F^{(0)}_{n,1}\ =\ \sum_{j=3}^{n} \sum_{i=3}^{j} R'_{1ij}\, +\, \sum_{j=5}^{n+1}\sum_{i=3}^{j-2}  R''_{1ij}
\ , 
\end{equation}
where the sum is performed with a periodic identification $n+1\!\sim\!1$. This expression was derived using a $[1 \, 2\rangle$ shift, and as a consequence   all of  the  $R$-invariants involved have $r\!=\!1$, and one can simply use the region momenta assignment shown in Figure \ref{exassign}. Using BCFW recursion relations it is possible to show that, for arbitrary helicity configuration, the tree-level form factor can be written in terms of $R'$ and $R''$. Therefore, one simply needs to show that these two functions are dual conformal invariant. 

It turns out that, for $s\!\neq\!t$, $R'$ and $R''$ are given by \eqref{Rfunction}, with the region variables assignment described below \eqref{supermomenta} (see also Figure \ref{exassign}). 
There is however a limiting case that needs to be discussed separately. For the specific configuration $R'_{rss}$,  \eqref{Rfunction} does not apply and one has instead
\begin{align}
R'_{rss} =
\raisebox{-.48\height}{
\begin{tikzpicture}[thick, scale=0.8]
  \drawLLwhite
  \drawUL{$\scriptstyle{0}$}
  \drawURempty
  \drawLR{$\scriptstyle{0}$}
  \drawboxinternallines
  \drawregionvariables{$x_c$}{$x_{c+1}$}{$x_a$}{$x_b$}{}
  \draw (UL) -- ++( 180:0.8) node[anchor=east] {$r+1$};
  \draw[dotted] (UL)+( 170:0.6) to [bend left=45] ++( 100:0.6);
  \draw (UL) -- ++(  90:0.8) node[anchor=south] {$s-1$};
  \draw[double] (UR) -- ++(  45:0.8) node {};
  \draw (LR) -- ++(   0:0.8) node[anchor=west] {$s$};
  \draw[dotted] (LR)+( -10:0.6) to [bend left=45] ++( -80:0.6);
  \draw (LR) -- ++( -90:0.8) node[anchor=north] {$r-1$}; 
  \draw (LL) -- ++(-135:0.8) node[anchor=north east] {$r$};
\end{tikzpicture}
} = -\frac{\langle s-1\,s\rangle \, \delta^{(4)}(\langle r|x_{ca}x_{ab}|\theta_{bc}\rangle  + \langle r|x_{cb}x_{ba}|\theta_{ac}\rangle )}{x_{ab}^4 \langle r|x_{cb}x_{ba}|s-1\rangle  \langle r|x_{ca}x_{ab}|s\rangle  \langle r|x_{ca}x_{bc}|r\rangle } \;. \label{Rfunregspec}
\end{align}
Notice that in this case $x_a=x_b^-$ and $x_{ab}=-q$. Taking the ratio with the limiting case of \eqref{Rfunction}, one can rewrite \eqref{Rfunregspec} as 
\begin{align}\label{Rsspref}
R'_{rss} &= -\frac{\langle r|x_{ca}x_{ab}|s-1\rangle  \langle r|x_{cb}x_{ba}|s\rangle}{x^2_{ab} \langle s-1\,s\rangle \langle r|x_{ca}x_{bc}|r\rangle} \; [(s-1)^-,s^-,s-1,s,r] \;.
\end{align}
As was shown in \cite{Bork:2014eqa}, the prefactor in \eqref{Rsspref} can be written as a ratio of four-brackets \eqref{fourbrack}. Since the four-bracket \eqref{fourbrack} is invariant under dual conformal transformations, once the prefactor is written in that form, we just need to check that it is also invariant under the little group scaling \eqref{scalingperiodic}. To this end, we first note that one can recast $R'_{rss}$ as
\begin{align}\label{Rssdci}
R'_{rss}=\frac{\braket{r,(s-1)^-,s^-,s-1} \braket{r,s-1,s,s^-}}{\braket{r^+, s-1, s, r} \braket{s, s^-, s-1, (s-1)^-} }\ [(s-1)^-,s^-,s-1,s,r] \ .
\end{align}
The novel feature of \eqref{Rssdci} is that the prefactor contains brackets involving one region variable as well as  its image after one period. To see how this happens consider the expression  $\langle r|x_{ca}x_{bc}|r\rangle $, which can be rewritten as
\begin{align}
  \langle r|x_{ca}x_{bc}|r\rangle = \langle r|(x_{c}^+-x_b)x_{bc}|r\rangle =\frac{\braket{r^+,s-1,s,r}}{\braket{s-1\, s}} \ .
\end{align}
Notice also that, by using a similar argument, it is easy to show that  the four-bracket is invariant under  an overall translation by a period:
\begin{align}
 \braket{r^+,s-1,s,r}=\braket{r,(s-1)^-,s^-,r^-} \ .
\end{align}
This is actually a trivial statement since we know that the four-bracket is invariant under the full dual conformal group and dual translations are just a subgroup.
Furthermore, since the little group transformation \eqref{scalingperiodic} does not depend on the specific period, we conclude that $R'_{rss}$ is invariant under little group scaling, and consequently  is a good dual conformal~invariant.

\section{Anomaly of one-loop form factors: a general proof}\label{oneloop}

In \cite{Brandhuber:2009kh} a deep connection between  IR divergences of one-loop scattering amplitudes and the dual conformal anomaly was established. The argument of \cite{Brandhuber:2009kh} is based on the  fact that only unitarity cuts in  two-particle channels contribute to the discontinuity of the IR-divergent part of an amplitude. Therefore, in the multiparticle case, the phase space integration can be performed  strictly  in four dimensions, and dual conformal symmetry of the discontinuity essentially descends from the covariance of the tree-level ingredients. A careful analysis shows that the invariance of the discontinuity is sufficient to prove that no multiparticle invariant can be present in the dual conformal anomaly, confirming the structure previously conjectured in \cite{Drummond:2008vq} (see \cite{Brandhuber:2009kh} for additional details of this derivation). 

\begin{figure}[htbp]
\centering
\begin{tikzpicture}[thick, scale=1.2]
\greyblob (blobL) at (-1, 0) {$\scriptstyle{k}$};
\greyblob (blobR) at (+1, 0) {$\scriptstyle{0}$};
\drawuppercut
\drawlowercut
\draw[double] (blobL) -- ++(+120:1);
\draw (blobL) -- ++( 180:1);
\draw[dotted] (blobL)+( 230:0.8) to [bend left=30] ++( 190:0.8);
\draw (blobL) -- ++(-120:1);
\draw (blobR) -- ++( 45:1) node[above right=-2pt] {$i$};
\draw (blobR) -- ++(-45:1) node[below right=-2pt] {$i+1$};
\node[] at (1.0, +1.0) {$x_{i}$};
\node[] at (1.0, -1.0) {$x_{i+2}$};
\node[] at (2.0,  0.0) {$x_{i+1}$};
\node[] at (0, 0) {$x_0$};
\end{tikzpicture}
\caption{The only two-particle cut contributing to the IR divergent part of the form factor as well as to the dual conformal anomaly.}
\label{2ptcutanomaly}
\end{figure}
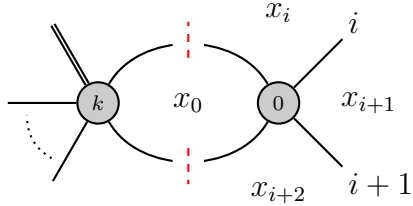

The argument can be extended to the case of form factors without any modification. Indeed, we know that the IR structure of the one-loop form factor is analogous to that of scattering amplitudes --  it depends only on two-particle invariants (see  \eqref{IRFF}). Therefore, the IR behaviour of one-loop form factors should be fully reproduced by the two-particle cut in Figure \ref{2ptcutanomaly}, which reads
\begin{align}
\left. F^{(1)}_{n,k} \right|_{x^2_{i,i+1} \text{cut}}\! =\int \mathrm{d}\text{LIPS}(\ell_1,\ell_2) \int \mathrm{d}^4 \eta_{\ell_1} \; \mathrm{d}^4 \eta_{\ell_2} \, A_{4,0}^{(0)}(i,i+1,\ell_2,\ell_1) F_{n,k}^{(0)}(-\ell_1,-\ell_2,i+2,\dots,i-1) \ .
\end{align}
The integration over fermionic variables can be immediately performed using the fermionic delta function of $A_{4,0}$, yielding
\begin{align}\label{cutIR}
 \left. F^{(1)}_{n,k} \right|_{x^2_{i,i+1} \text{cut}}\! =\int \mathrm{d}\text{LIPS}(\ell_1,\ell_2) \; \frac{\braket{\ell_1 \ell_2}^3 F_{n,k}^{(0)}(-\ell_1,-\ell_2,i+2,\dots,i-1)}{\braket{i,i+1}\braket{i+1,\ell_2}\braket{\ell_1,i}} \ .
\end{align}
Furthermore, using some spinor variable manipulations, we can rewrite \eqref{cutIR} as
\begin{align}\label{cutIR2}
 \left. F^{(1)}_{n,k} \right|_{x^2_{i,i+1} \text{cut}}\! =\int \mathrm{d}\text{LIPS}(\ell_1,\ell_2) \; \frac{\braket{\ell_1 \ell_2}^2 F_{n,k}^{(0)}(-\ell_1,-\ell_2,i+2,\dots,i-1)}{\braket{i,i+1}^2} \, \frac{x^2_{i,i+2}}{x^2_{0, i+1}} \ .
\end{align}
The crucial observation is that the IR-singular region of this integral is related to the collinear kinematic configuration
\begin{align}\label{kinconf}
 \ell_1&= -p_i \ , & \ell_2&=-p_{i+1} \, , & x_0&=x_{i+1} \ .
\end{align}
The divergence in the integral \eqref{cutIR2} is clearly related to the propagator $x_{0,i+1}^2$. The rest of the integrand can be evaluated in the configuration \eqref{kinconf}, and the cut can be uplifted to the corresponding integral, leading to
\begin{align}
 \left. F^{(1)}_{n,k} \right|_{\rm IR}=F_{n,k}^{(0)} \int \mathrm{d}^d x_0 \; \frac{x_{i,i+2}^2}{x_{0i}^2 x_{0,i+1}^2 x_{0,i+2}^2}
 \ , 
\end{align}
which evaluates to
\begin{align}\label{IRFF}
\left.F_{n,k}^{(1)}\right|_{\rm IR} = -F_{n,k}^{(0)}\, \sum_{i=1}^n \frac{( -x^2_{i i+2})^{-\eps}}{\eps^2} \;. 
\end{align}
This reproduces the correct IR behaviour of the form factor. 

The argument used in \cite{Brandhuber:2009kh} to relate the IR behaviour of scattering amplitudes to the expression of the dual conformal anomaly is based on the idea of applying a dual conformal transformation in the very first step of the above derivation, {\it i.e.}~on the two-particle cut. The covariance of the tree-level ingredients allows to show that the anomaly is related to the variation of the integration measure, which needs to be $d$-dimensional since the integral diverges (all the other two-particle cuts are finite and do not contribute to the anomaly). In particular, using the  definition of the  generator of dual special conformal transformations
\begin{equation}\label{Kmu}
  \mathsf{K}^{\mu}=\sum_{i=1}^n \left[-2 x_i^{\mu} x_i^{\nu} \frac{\partial}{\partial x_i^\nu} + x_i^2 \frac{\partial}{\partial x_i^\mu}\right] \ ,
\end{equation}
  the fact, proven in the previous section, that tree-level form factors transform covariantly,  and following steps similar to those of   \cite{Brandhuber:2009kh}, we arrive at 
 \begin{align}
\left. \mathsf{K}^{\mu}F^{(1)}_{n,k} \right|_{x^2_{i,i+1} \text{cut}}\! = 4\epsilon \int \mathrm{d}\text{LIPS}(\ell_1,\ell_2) 
\int \mathrm{d}^4 \eta_{\ell_1} \; \mathrm{d}^4 \eta_{\ell_2} \; x_0^{\mu}\; A_{4,0}^{(0)}(i,i+1,\ell_2,\ell_1) \; F_{n,k}^{(0)}(-\ell_1,-\ell_2,\dots)
\, , \end{align}
with $\eps=2-{d}/{2}$.
After this observation we can simply follow all the steps leading to \eqref{IRFF},  and hence 
we conclude that the one-loop anomaly has the form
\begin{align}
\label{anomaly-1}
\mathsf{K}^\mu \tilde{F}_{n,k}^{(1)} = -4 \, \tilde{F}_{n,k}^{(0)}\sum_{i=1}^n  \frac{x^\mu_{i+1}( -x^2_{i i+2})^{-\eps}}{\eps}\ . 
\end{align}
Note that the right-hand side of \eqref{anomaly-1} depends on the region momenta of the particles (and not just the momenta). 

Although the form of the anomaly resembles that of the amplitude case, the consequences for the one-loop expansion of the form factor in terms of scalar integrals are rather different. Indeed,  one-loop form factors may  contain three-mass triangles, which are finite in four dimensions and, in view of the previous arguments,  cannot contribute to the anomaly. On the other hand, we showed at the   beginning of this section that triangle integrals cannot be dual conformal invariant on their own. Therefore, two things can happen: either the variation of the finite triangles cancel some other variation arising  from the finite part of other integrals (in this case boxes); or the variation vanishes after summing over permutations. Notice also that, in the NMHV example, the three-mass triangle comes with a complicated coefficient, and  its variation needs to be taken into account as well  (see Section \ref{3mtriangle}).

To understand how the anomaly emerges, in the following we will  explicitly check its form  for MHV and NMHV form factors at one loop. Before doing that, we first elaborate on the consequences of   \eqref{anomaly-1} for the finite part of  one-loop form factors.
The universal IR-divergent part of a generic one-loop form factor has the form \eqref{IRFF}. Using 
\begin{align}
\label{basicx2}
\mathsf{K}^\mu x^2_{ab} = - 2 (x_a + x_b)^\mu \, x^2_{ab} 
\ , 
\end{align}
we can separate out the anomaly of the finite part. Doing so, one quickly arrives at
\begin{align}
\mathsf{K}^\mu \left.\tilde{F}_{n,k}^{(1)}\right|_{\rm fin} = -\tilde{F}_{n,k}^{(0)} \, \bigg[ \frac{2}{\eps} &\sum_{i=1}^n \Big( 2 x_{i+1}^\mu - (x_i^\mu + x_{i+2}^\mu)\Big)\, \cr 
- \,2 &\sum_{i=1}^n \Big( 2 x_{i+1}^\mu - (x_i^\mu + x_{i+2}^\mu)\Big) \log\left( - x^2_{i i+2} \right) 
\Bigg]
\ . 
\end{align}
The first sum evaluates to zero, thus we obtain
\begin{align}
\label{anom}
\mathsf{K}^\mu \left.\tilde{F}_{n,k}^{(1)}\right|_{\rm fin} = -2\, \tilde{F}_{n,k}^{(0)} \sum_{i=1}^n p_i^\mu \log\left( \frac{x^2_{i i+2}}{x^2_{i-1\,  i+1}}\right) 
\ , 
\end{align}
which, importantly, only depends on differences of region momenta ({\it i.e.}~momenta) and Mandelstam invariants of the particles. We now show the validity of this formula for the MHV and NMHV form factor at one loop.

\section{Examples}\label{examples}
Having presented a general derivation of the dual conformal anomaly, we now   analyse a number of  specific examples, namely  the one-loop MHV and NMHV form factors. The latter are particularly interesting due to the presence of a three-mass triangle, whose variation requires  a novel cancellation mechanism to be consistent with our general result \eqref{anomaly-1} and~\eqref{anom}.

There is an important preliminary observation to be made -- in order to find the correct anomaly, it is crucial to assign region variables according to the prescription  described in 
Section~\ref{tree_level} and illustrated in Figure~\ref{exassign}. In particular, this has to be done diagram by diagram in the expansion of the result in terms of scalar integrals;  crucially, the definition of the period $q$ in terms of region variables, and consequently its variation under special conformal transformations, is different for each of the diagrams involved in the computation. Let us now see how this works in  practice.

\subsection{$n$-point MHV form factor at one loop}

\begin{figure}[htb]
\centering
\begin{tikzpicture}[thick, scale=0.8]
  \drawLLempty;
  \drawULempty;
  \drawURempty;
  \drawLRempty;
  \drawboxinternallines
  \drawregionvariables{$x_1$}{$x_2$}{$x_i$}{$x_{i+1}$}{}
  \draw (UL) -- ++( 150:1.0);
  \draw[dotted] (UL)+( 145:0.8) to [bend left=30] ++( 125:0.8);
  \draw (UL) -- ++( 120:1.0);
  \draw (UR) -- ++(  45:1.0) node[above right=-2pt] {$i$};
  \draw (LR) -- ++( -15:1.0);
  \draw[dotted] (LR)+( -20:0.8) to [bend left=30] ++( -40:0.8);
  \draw (LR) -- ++( -45:1.0);
  \draw[double] (LR) -- ++(-75:1.0);
  \draw (LL) -- ++(-135:1.0) node[below left=-2pt] {$1$};
\end{tikzpicture}
\hspace{3cm}
\begin{tikzpicture}[thick, scale=0.8]
  \drawLLempty;
  \drawULempty;
  \drawURempty;
  \drawLRempty;
  \drawboxinternallines
  \drawregionvariables{$x_1^-$}{$x_2^-$}{$x_i$}{$\;x_{i+1}$}{}
  \draw (UL) -- ++( 165:1.0);
  \draw[dotted] (UL)+( 160:0.8) to [bend left=30] ++( 140:0.8);
  \draw (UL) -- ++( 135:1.0);
  \draw[double] (UL) -- ++( 105:1.0);
  \draw (UR) -- ++(  45:1.0) node[above right=-2pt] {$i$};
  \draw (LR) -- ++( -30:1.0);
  \draw[dotted] (LR)+( -35:0.8) to [bend left=30] ++( -55:0.8);
  \draw (LR) -- ++( -60:1.0);
  \draw (LL) -- ++(-135:1.0) node[below left=-2pt] {$1$};
\end{tikzpicture}
\caption{\it The two possible types of two-mass easy box functions. The double line  represents the incoming momentum $q$ of the operator. Note the two different assignments of region momenta in the two cases.}
\label{fig:all2me}
\end{figure}
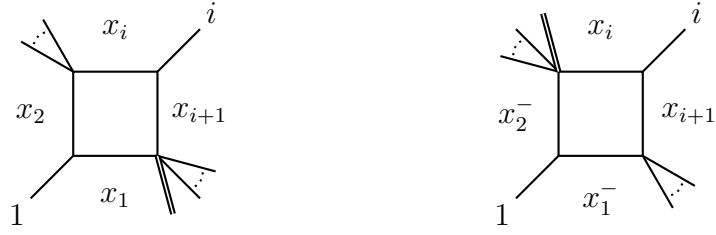

The generic one-loop MHV super form factor can be written compactly as \cite{Brandhuber:2010ad}: 
\begin{align}
\label{MHV-n}
F^{(1)}_{n,0} = F^{(0)}_{n,0} \left( -\sum_{i=1}^n \frac{(-x_{i\,i+2}^2)^{-\eps}}{\eps^2}\ + \ 
\sum_{r, a} \raisebox{-.47\height}{
\begin{tikzpicture}[thick, scale=0.65]
  \drawLLempty;
  \drawULempty;
  \drawURempty;
  \drawLRempty;
  \drawboxinternallines
  \drawregionvariables{$x_r$}{$x_{r+1}\;$}{$x_a$}{$\;x_{a+1}$}{F}
  \draw (UL) -- ++( 150:1.0);
  \draw[dotted] (UL)+( 145:0.8) to [bend left=30] ++( 125:0.8);
  \draw (UL) -- ++( 120:1.0);
  \draw (UR) -- ++(  45:1.0);
  \draw (LR) -- ++( -30:1.0);
  \draw[dotted] (LR)+( -35:0.8) to [bend left=30] ++( -55:0.8);
  \draw (LR) -- ++( -60:1.0);
  \draw (LL) -- ++(-135:1.0);
\end{tikzpicture}
} \right) \;. 
\end{align}
where the label F inside the box  indicates  the finite part of the reduced box integral \eqref{2measy}.
The sum is over all possible boxes; the off-shell leg can appear in  both  massive corners of the box function. The recipe to write the previous expression in terms of region variables depends as usual on the position of the off-shell legs, and an example is shown in Figure \ref{fig:all2me}. In that case the leg with momentum $p_1$ is associated to one of the massless legs and the region variables are assigned according to the two possible locations of the off-shell leg. A similar recipe can be applied to the other cyclic permutations.
 
In the following we will act with dual conformal generators on the finite part of a generic one-loop MHV super form factor.  
We will use the following two general formulae, obtained as repeated applications of \eqref{basicx2}: 
\begin{align}
\mathsf{K}^\mu \Li\left(1 - \frac{x_{ab}^2}{x_{ac}^2}\right) &= 2 x_{ab}^2 \, \frac{\log(x_{ab}^2 / x_{ac}^2)}{x_{ab}^2 - x_{ac}^2} \, x_{bc}^\mu \;, \\
\mathsf{K}^\mu \frac{1}{2} \log^2 \left(\frac{x_{ab}^2}{x_{a+1\, b+1}^2}\right) &= -2 \log \left(\frac{x_{ab}^2}{x_{a+1\, b+1}^2}\right) (x_{a\, a+1}^\mu + x_{b\, b+1}^\mu) \;.
\end{align}

Without loss of generality, we will now compute the term in the anomaly of the finite part of the $n$-point MHV form factor that is proportional to the momentum $p_1$. It is easy to realise that such terms can only arise from box functions where $p_1$ is one of the two massless legs. To perform the calculation we need to distinguish  terms where the form factor momentum is inserted in the two possible massive corners of a two-mass easy box. These two situations are depicted in Figure~\ref{fig:all2me}.  
The term proportional to $p_1$ in the variation of the first  type of box gives
	
\begin{align}
\mathsf{K}^{\mu} \raisebox{-.47\height}{\begin{tikzpicture}[thick, scale=0.5]
  \drawLLempty;
  \drawULempty;
  \drawURempty;
  \drawLRempty;
  \drawboxinternallines
  \drawregionvariables{$x_1$}{$x_2$}{$x_i$}{$\, \; x_{i+1}$}{F}
  \draw (UL) -- ++( 150:1.0);
  \draw[dotted] (UL)+( 145:0.8) to [bend left=30] ++( 125:0.8);
  \draw (UL) -- ++( 120:1.0);
  \draw (UR) -- ++(  45:1.0) node[above right=-2pt] {$i$};
  \draw (LR) -- ++( -15:1.0);
  \draw[dotted] (LR)+( -20:0.8) to [bend left=30] ++( -40:0.8);
  \draw (LR) -- ++( -45:1.0);
  \draw[double] (LR) -- ++(-75:1.0);
  \draw (LL) -- ++(-135:1.0) node[below left=-2pt] {$1$};
\end{tikzpicture}}
\begin{split}
\label{term1}
\sim-2 p_1^\mu\left[ \frac{x_{i+1\, 1}^2 }{ 	x_{i+1\, 1}^2 - x_{i+1\, 2}^2} \log \frac{x_{i+1\, 1}^2}{ x_{i+1\, 2}^2} \, - \, 
\frac{x_{2\, i}^2}{ x_{2\, i}^2 - x_{1\, i}^2 } \log\frac{x_{2\, i}^2 }{ x_{1\, i}^2}
\, - \,  \log\frac{x_{1\, i}^2 }{ x_{2\, i+1}^2}\right]
\ , 
\end{split}
\end{align}
	while for the second type of box we have
	\begin{align}
	\mathsf{K}^{\mu} \raisebox{-.47\height}{\begin{tikzpicture}[thick, scale=0.5]
  \drawLLempty;
  \drawULempty;
  \drawURempty;
  \drawLRempty;
  \drawboxinternallines
  \drawregionvariables{$x_1^-$}{$x_2^-$}{$x_i$}{$\; \, x_{i+1}$}{F}
  \draw (UL) -- ++( 165:1.0);
  \draw[dotted] (UL)+( 160:0.8) to [bend left=30] ++( 140:0.8);
  \draw (UL) -- ++( 135:1.0);
  \draw[double] (UL) -- ++( 105:1.0);
  \draw (UR) -- ++(  45:1.0) node[above right=-2pt] {$i$};
  \draw (LR) -- ++( -30:1.0);
  \draw[dotted] (LR)+( -35:0.8) to [bend left=30] ++( -55:0.8);
  \draw (LR) -- ++( -60:1.0);
  \draw (LL) -- ++(-135:1.0) node[below left=-2pt] {$1$};
\end{tikzpicture}}
\label{term2}
\sim -2 p_1^\mu&\left[ \frac{(x_{i+1,1}^+)^2}{(x^+_{i+1,1})^2 - (x_{i+1,2}^+)^2} \log \frac{(x_{i+1,1}^+)^2}{(x_{i+1,2}^+)^2}  - 
\frac{(x_{2 i}^-)^2}{(x_{2 i}^-)^2 - (x_{1 i}^-)^2 } \log\frac{(x_{2 i}^-)^2}{(x_{1 i}^-)^2}\right. \nonumber \\  &\left.-  \log\frac{(x_{1 i}^-)^2}{(x_{2 i+1}^-)^2}\right]
\ .
\end{align}
 Combining the variations and performing the sums
\begin{align}
\begin{split}
- \sum_{i=2}^{n-2}  \log \frac{ (x_{i+1, 1}^+)^2}{ (x_{i+1, 2}^+)^2} + 
\sum_{i=3}^{n-1} \log \frac{ x_{i+1, 1}^2}{ x_{i+1, 2}^2} 
+ \sum_{i=2}^{n-1} \log \frac{ (x_{1i}^-)^2}{ (x_{2, i+1}^-)^2}  + 
\sum_{i=3}^{n}  \log \frac{ x_{1 i}^2}{ x_{2, i+1}^2} \ , 
\end{split}
\end{align}
we obtain 
\begin{align}
\sum_{i} \mathsf{K}^{\mu} \Bigg( \raisebox{-.47\height}{\begin{tikzpicture}[thick, scale=0.5]
  \drawLLempty;
  \drawULempty;
  \drawURempty;
  \drawLRempty;
  \drawboxinternallines
  \drawregionvariables{$x_1$}{$x_2$}{$x_i$}{$\, \; x_{i+1}$}{F}
  \draw (UL) -- ++( 150:1.0);
  \draw[dotted] (UL)+( 145:0.8) to [bend left=30] ++( 125:0.8);
  \draw (UL) -- ++( 120:1.0);
  \draw (UR) -- ++(  45:1.0) node[above right=-2pt] {$i$};
  \draw (LR) -- ++( -15:1.0);
  \draw[dotted] (LR)+( -20:0.8) to [bend left=30] ++( -40:0.8);
  \draw (LR) -- ++( -45:1.0);
  \draw[double] (LR) -- ++(-75:1.0);
  \draw (LL) -- ++(-135:1.0) node[below left=-2pt] {$1$};
\end{tikzpicture}}+\raisebox{-.47\height}{\begin{tikzpicture}[thick, scale=0.5]
  \drawLLempty;
  \drawULempty;
  \drawURempty;
  \drawLRempty;
  \drawboxinternallines
  \drawregionvariables{$x_1^-$}{$x_2^-$}{$x_i$}{$\; \, x_{i+1}$}{F}
  \draw (UL) -- ++( 165:1.0);
  \draw[dotted] (UL)+( 160:0.8) to [bend left=30] ++( 140:0.8);
  \draw (UL) -- ++( 135:1.0);
  \draw[double] (UL) -- ++( 105:1.0);
  \draw (UR) -- ++(  45:1.0) node[above right=-2pt] {$i$};
  \draw (LR) -- ++( -30:1.0);
  \draw[dotted] (LR)+( -35:0.8) to [bend left=30] ++( -55:0.8);
  \draw (LR) -- ++( -60:1.0);
  \draw (LL) -- ++(-135:1.0) node[below left=-2pt] {$1$};
\end{tikzpicture}}\Bigg)\sim
-2 \, p_1^\mu \, \log \left( \frac{x_{1 \, 3}^2}{(x^{-}_{2\, n})^2} \right)
\ , 
\end{align}
in agreement with the term proportional to $p_1$ on the right-hand side of \eqref{anom}. 
Summarising, we have shown that the finite part of the dual conformal anomaly of an $n$-point MHV form factor is exactly reproduced by our general formula  \eqref{anom}. Next, we move on to consider NMHV form factors. 

\subsection{One-loop NMHV form factor}

The one-loop NMHV form factor can be computed using generalised unitarity as a combination of boxes and triangles \cite{Bork:2012tt}. The presence of the latter constitutes an important difference compared to amplitudes. 
In particular, for amplitudes the box integrals are invariant on their own%
\footnote{To be precise they are anomalous as we will discuss in Section \ref{boxes}.},  
and in addition  their coefficients are invariant as well.

For form factors  one may expect dual conformal symmetry to be broken. However, in the following  we will discover a new cancellation mechanism that ensures that the final result is invariant up to the expected anomaly.  The three-point NMHV form factor coincides with the $\overline{\text{MHV}}$ result,  and therefore  can be extracted from the MHV case considered earlier  by conjugation (this is analogous to the case of the five-point amplitude). 
The first  interesting case is that of a 
four-point NMHV form factor, as this is the first example which has a three-mass triangle. 
Since two-mass and one-mass triangles are IR divergent with vanishing finite part, their coefficient can be fixed by requiring a consistent divergent part for the final form factor, {\it i.e.}~\eqref{IRFF}.
On the other hand, the three-mass triangle is finite, and its coefficient has to be determined independently. 

We start by writing $\tilde{F}^{(1)}_{4,1}$ as a linear combination of reduced scalar integrals: 
\begin{align}
\tilde{F}^{(1)}_{4,1} &= b^{\text{1m}}
\raisebox{-.47\height}{
\begin{tikzpicture}[thick,scale=0.5]
\drawULempty
\drawURempty
\drawLRempty
\drawLLempty
\draw (UL) -- (UR) -- (LR) -- (LL) -- (UL);
\draw (UL) -- ++( 135:1.0) node[above left=-2pt] {$2$};
\draw (UR) -- ++(  45:1.0) node[above right=-2pt] {$3$};
\draw (LR) -- ++(   0:1.0) node[right=-2pt] {$4$};
\draw[double] (LR) -- ++( -90:1.0);
\draw (LL) -- ++(-135:1.0) node[below left=-2pt] {$1$};
\end{tikzpicture}
} + b_1^{\text{2mh}}
\raisebox{-.5\height}{
\begin{tikzpicture}[thick,scale=0.5]
\drawULempty
\drawURempty
\drawLRempty
\drawLLempty
\draw (UL) -- (UR) -- (LR) -- (LL) -- (UL);
\draw (UL) -- ++( 135:1.0) node[above left=-2pt] {$4$};
\draw[double] (UR) -- ++(  45:1.0);
\draw (LR) -- ++(   0:1.0) node[right=-2pt] {$1$};
\draw (LR) -- ++( -90:1.0) node[below=-2pt] {$2$};
\draw (LL) -- ++(-135:1.0) node[below left=-2pt] {$3$};
\end{tikzpicture}
} + b_2^{\text{2mh}}
\raisebox{-.45\height}{
\begin{tikzpicture}[thick,scale=0.5]
\drawULempty
\drawURempty
\drawLRempty
\drawLLempty
\draw (UL) -- (UR) -- (LR) -- (LL) -- (UL);
\draw (UL) -- ++( 135:1.0) node[above left=-2pt] {$1$};
\draw (UR) -- ++(  90:1.0) node[above=-2pt] {$2$};
\draw (UR) -- ++(   0:1.0) node[right=-2pt] {$3$};
\draw[double] (LR) -- ++( -45:1.0);
\draw (LL) -- ++(-135:1.0) node[below left=-2pt] {$4$};
\end{tikzpicture}
} \nonumber \\
&+c^{\text{2m}}
\raisebox{-.47\height}{
\begin{tikzpicture}[thick, scale=0.5]
\coordinate (T1) at ( 0.5 ,  0.87);
\coordinate (T2) at ( 0.5 , -0.87);
\coordinate (T3) at (-1.0 ,  0.0 );
\draw (T1) -- (T2) -- (T3) -- (T1);
\node[] (Tlbl0) at ( 0.0 ,  0.0) {};
\draw (T1) -- ++( 45:0.8) node[above right=-2pt] {$4$};
\draw (T2) -- ++(  0:0.8) node[right=-2pt] {$1$};
\draw (T2) -- ++(  -45:0.8) node[below right=-2pt] {$2$};
\draw (T2) -- ++(-90:0.8) node[below=-2pt] {$3$};
\draw[double] (T3) -- ++(180:0.8) node[anchor=east] {};
\end{tikzpicture}
}+c^{\text{3m}}
\raisebox{-.47\height}{
\begin{tikzpicture}[thick, scale=0.5]
\coordinate (T1) at ( 0.5 ,  0.87);
\coordinate (T2) at ( 0.5 , -0.87);
\coordinate (T3) at (-1.0 ,  0.0 );
\draw (T1) -- (T2) -- (T3) -- (T1);
\node[] (Tlbl0) at ( 0.0 ,  0.0) {};
\draw (T1) -- ++( 60:0.8) node[above right=-2pt] {$1$};
\draw (T1) -- ++(  0:0.8) node[right=-2pt] {$2$};
\draw (T2) -- ++(  0:0.8) node[right=-2pt] {$3$};
\draw (T2) -- ++(-60:0.8) node[below right=-2pt] {$4$};
\draw[double] (T3) -- ++(180:0.8) node[anchor=east] {};
\end{tikzpicture}
}+\text{cyclic}\, , 
\end{align}
where the sum is performed over cyclic permutations of the external legs. Notice that the dependence of the coefficients on the external momenta is understood and must be permuted accordingly. 
This is an expansion in terms of reduced scalar integrals, {\it i.e.}~where a dimensionful constant in the integral has been reabsorbed in the coefficient (see Appendix~\ref{integrals} for details). The coefficients of this linear combination have been determined in \cite{Bork:2012tt}. Here we review that derivation and  consider the transformation of the result under dual conformal symmetry. We start by the contribution of boxes and divergent triangles. 

\subsubsection{Boxes and divergent triangles}\label{boxes}
The contribution of boxes is easily computed using the maximal cuts. Each of the diagrams receives a contribution from two different cuts. In particular
\begin{align}
b^{\text{1m}}&=\frac12
\raisebox{-.48\height}{
\begin{tikzpicture}[thick, scale=0.8]
  \drawLLwhite
  \drawULblack
  \drawURwhite
  \drawLR{$\scriptstyle{1}$}
  \drawboxinternallines
  \draw (UL) -- ++( 135:0.8) node[above left=-2pt] {$2$};
  \draw (UR) -- ++(  45:0.8) node[above right=-2pt] {$3$};
  \draw (LR) -- ++(   0:0.8) node[right=-2pt] {$4$};
  \draw[double] (LR) -- ++( -90:0.8); 
  \draw (LL) -- ++(-135:0.8) node[below left=-2pt] {$1$};
\end{tikzpicture}
} +\frac12
\raisebox{-.48\height}{
\begin{tikzpicture}[thick, scale=0.8]
  \drawLLblack
  \drawULwhite
  \drawURblack
  \drawLR{$\scriptstyle{0}$}
  \drawboxinternallines
  \draw (UL) -- ++( 135:0.8) node[above left=-2pt] {$2$};
  \draw (UR) -- ++(  45:0.8) node[above right=-2pt] {$3$};
  \draw (LR) -- ++(   0:0.8) node[right=-2pt] {$4$};
  \draw[double] (LR) -- ++( -90:0.8); 
  \draw (LL) -- ++(-135:0.8) node[below left=-2pt] {$1$};
\end{tikzpicture}
}\\
b^{\text{2mh}}_1&=\frac12
\raisebox{-.50\height}{
\begin{tikzpicture}[thick, scale=0.8]
  \drawLLwhite
  \drawULblack
  \drawURempty
  \drawLR{$\scriptstyle{0}$}
  \drawboxinternallines
  \draw (UL) -- ++( 135:0.8) node[above left=-2pt] {$4$}; 
  \draw[double] (UR) -- ++(  45:0.8);
  \draw (LR) -- ++(   0:0.8) node[right=-2pt] {$1$};
  \draw (LR) -- ++( -90:0.8) node[below=-2pt] {$2$}; 
  \draw (LL) -- ++(-135:0.8) node[below left=-2pt] {$3$};
\end{tikzpicture}
} +\frac12 \raisebox{-.50\height}{
\begin{tikzpicture}[thick, scale=0.8]
  \drawLLblack
  \drawULwhite
  \drawURempty
  \drawLR{$\scriptstyle{0}$}
  \drawboxinternallines
  \draw (UL) -- ++( 135:0.8) node[above left=-2pt] {$4$}; 
  \draw[double] (UR) -- ++(  45:0.8);
  \draw (LR) -- ++(   0:0.8) node[right=-2pt] {$1$};
  \draw (LR) -- ++( -90:0.8) node[below=-2pt] {$2$}; 
  \draw (LL) -- ++(-135:0.8) node[below left=-2pt] {$3$};
\end{tikzpicture}
} \\
b^{\text{2mh}}_2&=\frac12
\raisebox{-.45\height}{
\begin{tikzpicture}[thick, scale=0.8]
  \drawLLblack
  \drawULwhite
  \drawUR{$\scriptstyle{0}$}
  \drawLRempty
  \drawboxinternallines
  \draw (UL) -- ++( 135:0.8) node[above left=-2pt] {$1$}; 
  \draw (UR) -- ++(   0:0.8) node[right=-2pt] {$3$};
  \draw (UR) -- ++(  90:0.8) node[above=-2pt] {$2$};
  \draw[double] (LR) -- ++( -45:0.8) node[below right=-2pt] {}; 
  \draw (LL) -- ++(-135:0.8) node[below left=-2pt] {$4$};
\end{tikzpicture}
} +\frac12\raisebox{-.45\height}{
\begin{tikzpicture}[thick, scale=0.8]
  \drawLLwhite
  \drawULblack
  \drawUR{$\scriptstyle{0}$}
  \drawLRempty
  \drawboxinternallines
  \draw (UL) -- ++( 135:0.8) node[above left=-2pt] {$1$}; 
  \draw (UR) -- ++(   0:0.8) node[right=-2pt] {$3$};
  \draw (UR) -- ++(  90:0.8) node[above=-2pt] {$2$};
  \draw[double] (LR) -- ++( -45:0.8) node[below right=-2pt] {}; 
  \draw (LL) -- ++(-135:0.8) node[below left=-2pt] {$4$};
\end{tikzpicture}
}
\end{align}
However, using the non-trivial identities \cite{Bork:2012tt, Part1}
\begin{align}\label{Rident1}
\raisebox{-.48\height}{
\begin{tikzpicture}[thick, scale=0.8]
  \drawLLwhite
  \drawULblack
  \drawURwhite
  \drawLR{$\scriptstyle{1}$}
  \drawboxinternallines
  \draw (UL) -- ++( 135:0.8) node[above left=-2pt] {$2$};
  \draw (UR) -- ++(  45:0.8) node[above right=-2pt] {$3$};
  \draw (LR) -- ++(   0:0.8) node[right=-2pt] {$4$};
  \draw[double] (LR) -- ++( -90:0.8); 
  \draw (LL) -- ++(-135:0.8) node[below left=-2pt] {$1$};
\end{tikzpicture}
} =
\raisebox{-.50\height}{
\begin{tikzpicture}[thick, scale=0.8]
  \drawLLwhite
  \drawULblack
  \drawURempty
  \drawLR{$\scriptstyle{0}$}
  \drawboxinternallines
  \draw (UL) -- ++( 135:0.8) node[above left=-2pt] {$4$}; 
  \draw[double] (UR) -- ++(  45:0.8);
  \draw (LR) -- ++(   0:0.8) node[right=-2pt] {$1$};
  \draw (LR) -- ++( -90:0.8) node[below=-2pt] {$2$}; 
  \draw (LL) -- ++(-135:0.8) node[below left=-2pt] {$3$};
\end{tikzpicture}
} =
\raisebox{-.45\height}{
\begin{tikzpicture}[thick, scale=0.8]
  \drawLLblack
  \drawULwhite
  \drawUR{$\scriptstyle{0}$}
  \drawLRempty
  \drawboxinternallines
  \draw (UL) -- ++( 135:0.8) node[above left=-2pt] {$1$}; 
  \draw (UR) -- ++(   0:0.8) node[right=-2pt] {$3$};
  \draw (UR) -- ++(  90:0.8) node[above=-2pt] {$2$};
  \draw[double] (LR) -- ++( -45:0.8) node[below right=-2pt] {}; 
  \draw (LL) -- ++(-135:0.8) node[below left=-2pt] {$4$};
\end{tikzpicture}
} = R'_{144} = R'_{311}\, , 
\end{align}
and
\begin{align}\label{Rident2}
\raisebox{-.48\height}{
\begin{tikzpicture}[thick, scale=0.8]
  \drawLLblack
  \drawULwhite
  \drawURblack
  \drawLR{$\scriptstyle{0}$}
  \drawboxinternallines
  \draw (UL) -- ++( 135:0.8) node[above left=-2pt] {$2$};
  \draw (UR) -- ++(  45:0.8) node[above right=-2pt] {$3$};
  \draw (LR) -- ++(   0:0.8) node[right=-2pt] {$4$};
  \draw[double] (LR) -- ++( -90:0.8); 
  \draw (LL) -- ++(-135:0.8) node[below left=-2pt] {$1$};
\end{tikzpicture}
} =
\raisebox{-.50\height}{
\begin{tikzpicture}[thick, scale=0.8]
  \drawLLblack
  \drawULwhite
  \drawURempty
  \drawLR{$\scriptstyle{0}$}
  \drawboxinternallines
  \draw (UL) -- ++( 135:0.8) node[above left=-2pt] {$4$}; 
  \draw[double] (UR) -- ++(  45:0.8);
  \draw (LR) -- ++(   0:0.8) node[right=-2pt] {$1$};
  \draw (LR) -- ++( -90:0.8) node[below=-2pt] {$2$}; 
  \draw (LL) -- ++(-135:0.8) node[below left=-2pt] {$3$};  
\end{tikzpicture}
} =
\raisebox{-.45\height}{
\begin{tikzpicture}[thick, scale=0.8]
  \drawLLwhite
  \drawULblack
  \drawUR{$\scriptstyle{0}$}
  \drawLRempty
  \drawboxinternallines
  \draw (UL) -- ++( 135:0.8) node[above left=-2pt] {$1$}; 
  \draw (UR) -- ++(   0:0.8) node[right=-2pt] {$3$};
  \draw (UR) -- ++(  90:0.8) node[above=-2pt] {$2$};
  \draw[double] (LR) -- ++( -45:0.8) node[below right=-2pt] {}; 
  \draw (LL) -- ++(-135:0.8) node[below left=-2pt] {$4$};
\end{tikzpicture}
} = R'_{241} = R''_{424} \;, 
\end{align}
and noticing that, by IR consistency, $c^{\text{2m}}$ is fixed to 
\begin{align}\label{c2m}
 c^{\text{2m}}=R'_{144}+R'_{241}\, , 
\end{align}
we arrive at the following compact expression for the NMHV four-point form factor:
\begin{align}\label{NMHV4pt1loop}
\tilde{F}^{(1)}_{4,1} &=\frac{c^{\text{2m}}}{2} \left(
\raisebox{-.47\height}{
\begin{tikzpicture}[thick,scale=0.5]
\drawULempty
\drawURempty
\drawLRempty
\drawLLempty
\draw (UL) -- (UR) -- (LR) -- (LL) -- (UL);
  \draw (UL) -- ++( 135:1.0) node[above left=-2pt] {$2$};
  \draw (UR) -- ++(  45:1.0) node[above right=-2pt] {$3$};
  \draw (LR) -- ++(   0:1.0) node[right=-2pt] {$4$};
  \draw[double] (LR) -- ++( -90:1.0); 
  \draw (LL) -- ++(-135:1.0) node[below left=-2pt] {$1$};
\end{tikzpicture}
} +
\raisebox{-.5\height}{
\begin{tikzpicture}[thick,scale=0.5]
\drawULempty
\drawURempty
\drawLRempty
\drawLLempty
\draw (UL) -- (UR) -- (LR) -- (LL) -- (UL);
  \draw (UL) -- ++( 135:1.0) node[above left=-2pt] {$4$}; 
  \draw[double] (UR) -- ++(  45:1.0);
  \draw (LR) -- ++(   0:1.0) node[right=-2pt] {$1$};
  \draw (LR) -- ++( -90:1.0) node[below=-2pt] {$2$}; 
  \draw (LL) -- ++(-135:1.0) node[below left=-2pt] {$3$};
\end{tikzpicture}
} +
\raisebox{-.45\height}{
\begin{tikzpicture}[thick,scale=0.5]
\drawULempty
\drawURempty
\drawLRempty
\drawLLempty
\draw (UL) -- (UR) -- (LR) -- (LL) -- (UL);
  \draw (UL) -- ++( 135:1.0) node[above left=-2pt] {$1$}; 
  \draw (UR) -- ++(   0:1.0) node[right=-2pt] {$3$};
  \draw (UR) -- ++(  90:1.0) node[above=-2pt] {$2$};
  \draw[double] (LR) -- ++( -45:1.0); 
  \draw (LL) -- ++(-135:1.0) node[below left=-2pt] {$4$};
\end{tikzpicture}
} 
+ \, 2
\raisebox{-.47\height}{
\begin{tikzpicture}[thick, scale=0.5]
\coordinate (T1) at ( 0.5 ,  0.87);
\coordinate (T2) at ( 0.5 , -0.87);
\coordinate (T3) at (-1.0 ,  0.0 );
\draw (T1) -- (T2) -- (T3) -- (T1);
\node[] (Tlbl0) at ( 0.0 ,  0.0) {};
\draw (T1) -- ++( 45:0.8) node[above right=-2pt] {$4$};
\draw (T2) -- ++(  0:0.8) node[right=-2pt] {$1$};
\draw (T2) -- ++(  -45:0.8) node[below right=-2pt] {$2$};
\draw (T2) -- ++(-90:0.8) node[below=-2pt] {$3$};
\draw[double] (T3) -- ++(180:0.8) node[left=-2pt] {};
\end{tikzpicture}
}\right)\nonumber 
\\&+c^{\text{3m}}
\raisebox{-.47\height}{
\begin{tikzpicture}[thick, scale=0.5]
\coordinate (T1) at ( 0.5 ,  0.87);
\coordinate (T2) at ( 0.5 , -0.87);
\coordinate (T3) at (-1.0 ,  0.0 );
\draw (T1) -- (T2) -- (T3) -- (T1);
\node[] (Tlbl0) at ( 0.0 ,  0.0) {};
\draw (T1) -- ++( 60:0.8) node[above right=-2pt] {$1$};
\draw (T1) -- ++(  0:0.8) node[right=-2pt] {$2$};
\draw (T2) -- ++(  0:0.8) node[right=-2pt] {$3$};
\draw (T2) -- ++(-60:0.8) node[below right=-2pt] {$4$};
\draw[double] (T3) -- ++(180:0.8);
\end{tikzpicture}
}+\text{cyclic}\, .
\end{align}
We focus here on the first line of \eqref{NMHV4pt1loop}, and  compute its variation under dual conformal transformations, while the three-mass triangle is discussed in Section \ref{3mtriangle}. The overall coefficient $c^{\text{2m}}$ is expressed in terms of $R$-invariants (see \eqref{c2m}) and therefore  is explicitly dual conformal invariant as  shown in Section \ref{tree_level}. Furthermore, in light of \eqref{anom}, we are interested in the finite part of the result and we can neglect the two-mass triangles, which are purely divergent. We then look at the particular combination
\begin{align}\label{Vdef}
V =
\raisebox{-.47\height}{
\begin{tikzpicture}[thick,scale=0.5]
\drawULempty
\drawURempty
\drawLRempty
\drawLLempty
 \drawregionvariables{}{}{}{}{F}
\draw (UL) -- (UR) -- (LR) -- (LL) -- (UL);
  \draw (UL) -- ++( 135:1.0) node[above left=-2pt] {$2$};
  \draw (UR) -- ++(  45:1.0) node[above right=-2pt] {$3$};
  \draw (LR) -- ++(   0:1.0) node[right=-2pt] {$4$};
  \draw[double] (LR) -- ++( -90:1.0); 
  \draw (LL) -- ++(-135:1.0) node[below left=-2pt] {$1$};
\end{tikzpicture}
} +
\raisebox{-.5\height}{
\begin{tikzpicture}[thick,scale=0.5]
\drawULempty
\drawURempty
\drawLRempty
\drawLLempty
 \drawregionvariables{}{}{}{}{F}
\draw (UL) -- (UR) -- (LR) -- (LL) -- (UL);
  \draw (UL) -- ++( 135:1.0) node[above left=-2pt] {$4$}; 
  \draw[double] (UR) -- ++(  45:1.0);
  \draw (LR) -- ++(   0:1.0) node[right=-2pt] {$1$};
  \draw (LR) -- ++( -90:1.0) node[below=-2pt] {$2$}; 
  \draw (LL) -- ++(-135:1.0) node[below left=-2pt] {$3$};
\end{tikzpicture}
} +
\raisebox{-.45\height}{
\begin{tikzpicture}[thick,scale=0.5]
\drawULempty
\drawURempty
\drawLRempty
\drawLLempty
 \drawregionvariables{}{}{}{}{F}
\draw (UL) -- (UR) -- (LR) -- (LL) -- (UL);
  \draw (UL) -- ++( 135:1.0) node[above left=-2pt] {$1$}; 
  \draw (UR) -- ++(   0:1.0) node[right=-2pt] {$3$};
  \draw (UR) -- ++(  90:1.0) node[above=-2pt] {$2$};
  \draw[double] (LR) -- ++( -45:1.0); 
  \draw (LL) -- ++(-135:1.0) node[below left=-2pt] {$4$};
\end{tikzpicture}
} \;,
\end{align}
where again the letter F indicates the finite part of the integral.

The variation of the scalar box integrals can be computed in two different ways: either one takes the variation of the integrands and then uses some reduction techniques to recast the result in terms of scalar triangles as was done in \cite{Brandhuber:2009xz}, or one just takes the variation of the finite part of the integrated result (explicit expressions can be found in Appendix~\ref{integrals}). Either way, the result is
\begin{align}
\mathsf{K}^\mu
\raisebox{-.47\height}{
\begin{tikzpicture}[thick,scale=0.5]
\drawULempty
\drawURempty
\drawLRempty
\drawLLempty
 \drawregionvariables{$x_1$}{$x_2$}{$x_3$}{$x_4$}{F}
\draw (UL) -- (UR) -- (LR) -- (LL) -- (UL);
  \draw (UL) -- ++( 135:1.0) node[above left=-2pt] {$2$};
  \draw (UR) -- ++(  45:1.0) node[above right=-2pt] {$3$};
  \draw (LR) -- ++(   0:1.0) node[right=-2pt] {$4$};
  \draw[double] (LR) -- ++( -90:1.0); 
  \draw (LL) -- ++(-135:1.0) node[below left=-2pt] {$1$};
\end{tikzpicture}
} = &\quad 2p_1^\mu \left(\frac{x_{14}^2}{x_{14}^2-x_{24}^2}\,\log\frac{x_{14}^2}{x_{13}^2} + \frac{x_{24}^2}{x_{14}^2-x_{24}^2}\,\log\frac{x_{13}^2}{x_{24}^2}\right) \cr
&\quad + 2p_3^\mu \left(\frac{x_{13}^2}{x_{13}^2-x_{14}^2}\,\log\frac{x_{24}^2}{x_{13}^2} + \frac{x_{14}^2}{x_{13}^2-x_{14}^2}\,\log\frac{x_{14}^2}{x_{24}^2}\right) \;, \\
\mathsf{K}^\mu
\raisebox{-.5\height}{
\begin{tikzpicture}[thick,scale=0.5]
\drawULempty
\drawURempty
\drawLRempty
\drawLLempty
 \drawregionvariables{$x_3$}{$x_4$}{$x_1^-$}{$x_1$}{F}
\draw (UL) -- (UR) -- (LR) -- (LL) -- (UL);
  \draw (UL) -- ++( 135:1.0) node[above left=-2pt] {$4$}; 
  \draw[double] (UR) -- ++(  45:1.0);
  \draw (LR) -- ++(   0:1.0) node[right=-2pt] {$1$};
  \draw (LR) -- ++( -90:1.0) node[below=-2pt] {$2$}; 
  \draw (LL) -- ++(-135:1.0) node[below left=-2pt] {$3$};
\end{tikzpicture}
} = &\quad -(p_1^\mu+p_2^\mu)\log\frac{(x_{13}^-)^2}{x_{13}^2} - q^\mu\log\frac{(x_{13}^-)^2}{q^2} + 2(p_1^\mu+p_2^\mu+p_4^\mu)\log\frac{(x_{13}^-)^2}{x_{14}^2} \cr
&\quad + 2p_3^\mu \, \frac{x_{13}^2}{x_{13}^2-x_{14}^2}\,\log\frac{x_{13}^2}{x_{14}^2} - 2p_4^\mu \, \frac{q^2}{q^2-x_{14}^2}\,\log\frac{q^2}{x_{14}^2} \;, \\
\mathsf{K}^\mu
\raisebox{-.45\height}{
\begin{tikzpicture}[thick,scale=0.5]
\drawULempty
\drawURempty
\drawLRempty
\drawLLempty
 \drawregionvariables{$x_4$}{$x_1^-$}{$x_2^-$}{$x_4^-$}{F}
\draw (UL) -- (UR) -- (LR) -- (LL) -- (UL);
  \draw (UL) -- ++( 135:1.0) node[above left=-2pt] {$1$}; 
  \draw (UR) -- ++(   0:1.0) node[right=-2pt] {$3$};
  \draw (UR) -- ++(  90:1.0) node[above=-2pt] {$2$};
  \draw[double] (LR) -- ++( -45:1.0); 
  \draw (LL) -- ++(-135:1.0) node[below left=-2pt] {$4$};
\end{tikzpicture}
} = &\quad (p_2^\mu+p_3^\mu)\log\frac{(x_{24}^-)^2}{x_{24}^2} + q^\mu\log\frac{(x_{24}^-)^2}{q^2} - 2(p_2^\mu+p_3^\mu+p_4^\mu)\log\frac{(x_{24}^-)^2}{x_{14}^2} \cr
&\quad - 2p_1^\mu \, \frac{x_{24}^2}{x_{24}^2-x_{14}^2}\,\log\frac{x_{24}^2}{x_{14}^2} + 2p_4^\mu \, \frac{q^2}{q^2-x_{14}^2}\,\log\frac{q^2}{x_{14}^2} \;.
\end{align}
Notice that, in computing these variations, the correct assignment of region variables is essential. As in our previous examples, we start assigning region variables from the position of the off-shell leg and then  follow the ordering along the periodic configuration. The variations above are then obtained  by writing each  integral using   their particular region variable assignment, and acting with the  generator $\mathsf{K}^{\mu}$ in \eqref{Kmu}. 
For the particular combination in \eqref{Vdef}, this gives
\begin{align}\label{KV}
\mathsf{K}^\mu V = p_1^\mu \log\frac{(x_{24}^-)^2}{x_{13}^2} + p_2^\mu \log\frac{x_{13}^2}{x_{24}^2} + p_3^\mu \log\frac{x_{24}^2}{(x_{13}^-)^2} + p_4^\mu \log\frac{(x_{13}^-)^2}{(x_{24}^-)^2} \;.
\end{align}
This surprisingly simple combination is invariant under cyclic permutations. Therefore, using  \eqref{NMHV4pt1loop} we can write
\begin{align}
\left.\mathsf{K}^\mu \tilde{F}^{(1)}_{4,1}\right|_{\text{fin}}= \frac{1}{2} \, \mathsf{K}^\mu V \sum_{\text{cyclic}} c^{\text{2m}} + \mathsf{K}^\mu T^{\text{3m}} \ , 
\end{align}
where $T^{\text{3m}}$ is the contribution of the three-mass triangles
\begin{align}\label{T3m}
 T^{\text{3m}}=c^{\text{3m}} \raisebox{-.47\height}{
\begin{tikzpicture}[thick, scale=0.5]
\coordinate (T1) at ( 0.5 ,  0.87);
\coordinate (T2) at ( 0.5 , -0.87);
\coordinate (T3) at (-1.0 ,  0.0 );
\draw (T1) -- (T2) -- (T3) -- (T1);
\node[] (Tlbl0) at ( 0.0 ,  0.0) {};
\draw (T1) -- ++( 60:0.8) node[above right=-2pt] {$1$};
\draw (T1) -- ++(  0:0.8) node[right=-2pt] {$2$};
\draw (T2) -- ++(  0:0.8) node[right=-2pt] {$3$};
\draw (T2) -- ++(-60:0.8) node[below right=-2pt] {$4$};
\draw[double] (T3) -- ++(180:0.8) node[anchor=east] {};
\end{tikzpicture}
}+\text{cyclic} \, .
\end{align}
The sum over cyclic permutations of $c^{\text{2m}}$ reads
\begin{align}
\sum_{\text{cyclic}} c^{{2m}} = R'_{144} + R'_{241} + R'_{211} + R'_{312} + R'_{322} + R'_{423} + R'_{433} + R'_{134} = 4\tilde{F}^{(0)}_{4,1}\, ,
\end{align}
where for the last equality we used  \eqref{genNHMV} combined with the identities \eqref{Rident1}, \eqref{Rident2} and permutations thereof. Expressing \eqref{KV} in terms of region variables we have
\begin{align}
\left.\mathsf{K}^\mu \tilde{F}^{(1)}_{4,1}\right|_{\text{fin}}=  -2\,  \tilde F_{4,1}^{(0)}  \sum_{i=1}^4 p_i^\mu \log\left( \frac{x^2_{i i+2}}{x^2_{i-1\,  i+1}}\right)  + \mathsf{K}^\mu T^{\text{3m}} \, .
\end{align}
This result implies that the boxes already account for the full anomaly \eqref{anom}. As a consequence, the necessary and sufficient condition for dual conformal invariance is 
\begin{align}
 \mathsf{K}^\mu T^{\text{3m}}=0 \ .
\end{align}
We will check this surprising relation   in the next section.

\subsubsection{Three-mass triangles}\label{3mtriangle}
In this section we show that the contribution of the three-mass triangles is dual conformal invariant. We start by reviewing the computation of $c^{\text{3m}}$. This coefficient is harder than the boxes' since it requires looking at non-maximal cuts. Nevertheless, a prescription  for the direct extraction of this coefficient was given in \cite{Forde:2007mi} and applied to the case of form factors in \cite{Bork:2012tt}. Let us consider the general configuration
\begin{align}\label{3mscalartriangle}
\raisebox{-.47\height}{
\begin{tikzpicture}[thick, scale=0.8]
\coordinate (T1) at ( 0.5 ,  0.87);
\coordinate (T2) at ( 0.5 , -0.87);
\coordinate (T3) at (-1.0 ,  0.0 );
\draw (T1) -- (T2) -- (T3) -- (T1);
\node[] (Tlbl3) at (-0.6 ,  0.8) {$x_a$};
\node[] (Tlbl1) at ( 1.2 ,  0.0) {$x_b$};
\node[] (Tlbl2) at (-0.6 , -0.8) {$x_c$};
\node[] (Tlbl0) at ( 0.0 ,  0.0) {};
\draw (T1) -- ++( 60:0.8) node[above right=-2pt] {$r$};
\draw[dotted] (T1)+( 50:0.6) to [bend left=45] ++( 10:0.6);
\draw (T1) -- ++(  0:0.8) node[right] {$s-1$};
\draw (T2) -- ++(  0:0.8) node[right] {$s$};
\draw[dotted] (T2)+(-10:0.6) to [bend left=45] ++(-50:0.6);
\draw (T2) -- ++(-60:0.8) node[below right=-2pt] {$r-1$};
\draw[double] (T3) -- ++(180:0.8);
\end{tikzpicture}
}
\end{align}
which contains an arbitrary number of legs, but no external momentum in the massive corner containing the off-shell leg. In \cite{Bork:2012tt} it was shown that only this type of diagrams arise in the computation of the one-loop NMHV form factor. Here, we will show that this structure is crucial for the dual conformal invariance of the coefficient $c^{\text{3m}}$, which would be spoiled by the presence of an external leg in the same corner of the off-shell leg. The four-point case can be immediately recovered by setting $r=1$ and $s=3$. Notice also that, for this particular configuration, $x_c=x_a^-$.

The starting point for the computation of $c^{\text{3m}}$ is the triple cut
\begin{align}\label{triplecut}
\raisebox{-.49\height}{
\begin{tikzpicture}[thick, scale=0.85]
\node[circle, fill=black!20, draw, minimum size=0.5cm, inner sep=2pt] (T1) at ( 0.5 ,  0.87) {$\scriptstyle{0}$};
\node[circle, fill=black!20, draw, minimum size=0.5cm, inner sep=2pt] (T2) at ( 0.5 , -0.87) {$\scriptstyle{0}$};
\coordinate (T3) at (-1.0 ,  0.0 );
\draw (T1) --node[circle, fill=white, draw=white]{} (T2) --node[circle, fill=white, draw=white]{} (T3) --node[circle, fill=white, draw=white]{} (T1);
\node[] (Tlbl3) at (-0.6 ,  0.8) {$x_a$};
\node[] (Tlbl1) at ( 1.2 ,  0.0) {$x_b$};
\node[] (Tlbl2) at (-0.6 , -0.8) {$x_c$};
\node[] (Tlbl0) at ( 0.0 ,  0.0) {$x_0$};
\draw[red, dashed] (0,0)+(  0:0.4) -- ++(  0:0.8);
\draw[red, dashed] (-0.15,0)+(120:0.3) -- ++(120:0.7);
\draw[red, dashed] (-0.15,0)+(240:0.3) -- ++(240:0.7);
\draw (T1) -- ++( 60:0.8) node[above right=-2pt] {$r$};
\draw[dotted] (T1)+( 50:0.6) to [bend left=45] ++( 10:0.6);
\draw (T1) -- ++(  0:0.8) node[right] {$s-1$};
\draw (T2) -- ++(  0:0.8) node[right] {$s$};
\draw[dotted] (T2)+(-10:0.6) to [bend left=45] ++(-50:0.6);
\draw (T2) -- ++(-60:0.8) node[below right=-2pt] {$r-1$};
\draw[double] (T3) -- ++(180:0.8);
\end{tikzpicture}} \!\!\!
= \int \prod_{i=1}^3 \mathrm{d}^4 \eta_{\ell_i} \; F^{(0)}_{2,0}(-\ell_3,\ell_1) \; A^{(0)}_{n_1,0}(-\ell_1,\dots,\ell_2)  \; A^{(0)}_{n_2,0}(-\ell_2,\dots,\ell_3) 
\end{align}
with
\begin{align}
\ell_1 = x_{a0} \;, \qquad \ell_2 = x_{b0} \;, \qquad \ell_3 = x_{c0} \;.
\end{align}
The integration over the fermionic variables yields
\begin{align}
 &\int \prod_{i=1}^3  \mathrm{d}^4  \eta_{\ell_i} \; \d^8(\eta_{\ell_1} \lambda_{\ell_1}-\eta_{\ell_3} \lambda_{\ell_3}+\theta_{ca}) \; \d^8(\eta_{\ell_2} \lambda_{\ell_2}-\eta_{\ell_1} \lambda_{\ell_1}+\theta_{ab}) \; \d^8(\eta_{\ell_3} \lambda_{\ell_3}-\eta_{\ell_2} \lambda_{\ell_2}+\theta_{bc}) \nonumber\\
 &=\d^8(\mathrm{q}_{\text{tot}})\int \prod_{i=1}^3 \mathrm{d}^4  \eta_{\ell_i} \; \d^{(4)}(\braket{\ell_1 \, \ell_2}\eta_{\ell_2} +\braket{\ell_1 \, \theta_{ab}}) \; \d^{(4)}(\braket{\ell_1 \, \ell_2}\eta_{\ell_1} +\braket{\ell_2 \, \theta_{ab}}) \; \frac{1}{\braket{\ell_1 \, \ell_2}^4} \nonumber\\
 &\phantom{\d^8(\mathrm{q}_{\text{tot}})\int \prod_{i=1}^3  d^4  \eta_{\ell_i}} \times \;\; \d^{(4)}(\braket{\ell_2 \, \ell_3}\eta_{\ell_3} +\braket{\ell_2 \, \theta_{bc}}) \; \d^{(4)}(\braket{\ell_2 \, \ell_3}\eta_{\ell_2} +\braket{\ell_3 \, \theta_{bc}}) \; \frac{1}{\braket{\ell_2 \, \ell_3}^4} \nonumber\\
 &=\d^8(\mathrm{q}_{\text{tot}}) \; \d^{(4)}(\braket{\ell_1 \, \ell_2}\braket{\ell_3 \, \theta_{bc}} -\braket{\ell_2 \, \ell_3}\braket{\ell_1 \, \theta_{ab}}) \;.
\end{align}
After these manipulations the three-particle cut reads
\begin{align}
\raisebox{-.49\height}{
\begin{tikzpicture}[thick, scale=0.85]
\node[circle, fill=black!20, draw, minimum size=0.5cm, inner sep=2pt] (T1) at ( 0.5 ,  0.87) {$\scriptstyle{0}$};
\node[circle, fill=black!20, draw, minimum size=0.5cm, inner sep=2pt] (T2) at ( 0.5 , -0.87) {$\scriptstyle{0}$};
\coordinate (T3) at (-1.0 ,  0.0 );
\draw (T1) --node[circle, fill=white, draw=white]{} (T2) --node[circle, fill=white, draw=white]{} (T3) --node[circle, fill=white, draw=white]{} (T1);
\node[] (Tlbl3) at (-0.6 ,  0.8) {$x_a$};
\node[] (Tlbl1) at ( 1.2 ,  0.0) {$x_b$};
\node[] (Tlbl2) at (-0.6 , -0.8) {$x_c$};
\node[] (Tlbl0) at ( 0.0 ,  0.0) {$x_0$};
\draw[red, dashed] (0,0)+(  0:0.4) -- ++(  0:0.8);
\draw[red, dashed] (-0.15,0)+(120:0.3) -- ++(120:0.7);
\draw[red, dashed] (-0.15,0)+(240:0.3) -- ++(240:0.7);
\draw (T1) -- ++( 60:0.8) node[above right=-2pt] {$r$};
\draw[dotted] (T1)+( 50:0.6) to [bend left=45] ++( 10:0.6);
\draw (T1) -- ++(  0:0.8) node[right] {$s-1$};
\draw (T2) -- ++(  0:0.8) node[right] {$s$};
\draw[dotted] (T2)+(-10:0.6) to [bend left=45] ++(-50:0.6);
\draw (T2) -- ++(-60:0.8) node[below right=-2pt] {$r-1$};
\draw[double] (T3) -- ++(180:0.8);
\end{tikzpicture}} \!\!\!
= F^{(0)}_{n,0} \, \frac{\langle s-1\,s\rangle \langle r-1\,r\rangle \,\d^{(4)}(\braket{\ell_1 \, \ell_2}\braket{\ell_3 \, \theta_{bc}} -\braket{\ell_2 \, \ell_3}\braket{\ell_1 \, \theta_{ab}}) }{\langle r\,\ell_1\rangle \langle s-1\,\ell_2\rangle \langle s\,\ell_2\rangle \langle r-1\,\ell_3\rangle \langle \ell_1\,\ell_2\rangle \langle\ell_2\,\ell_3\rangle \langle\ell_1\,\ell_3\rangle ^2} \;,  \label{triplecut2}
\end{align}
and the associated coefficient is
\begin{align}\label{c3mdci}
 c^{\text{3m}}=\frac{\langle s-1\,s\rangle \langle r-1\,r\rangle \,\d^{(4)}(\braket{\ell_1 \, \ell_2}\braket{\ell_3 \, \theta_{bc}} -\braket{\ell_2 \, \ell_3}\braket{\ell_1 \, \theta_{ab}}) }{\Delta_{abc} \langle r\,\ell_1\rangle \langle s-1\,\ell_2\rangle \langle s\,\ell_2\rangle \langle r-1\,\ell_3\rangle \langle \ell_1\,\ell_2\rangle \langle\ell_2\,\ell_3\rangle \langle\ell_1\,\ell_3\rangle ^2} \, , 
\end{align}
where $\Delta_{abc}$ is defined in \eqref{Delta3m} and  originates from expanding the form factor  in a basis of reduced triangles, see  \eqref{reducedtriangles}. A similar factor would appear for the case of boxes, but it always cancels after evaluating the quadruple cut on the corresponding solution. Here a similar cancellation does not seem to happen and we will have to deal with this additional factor. Furthermore, the MHV factor in \eqref{triplecut2} has been removed because the expansion \eqref{T3m} refers to the ratio $\tilde F^{(0)}_{4,1}$. 

As usual, in \eqref{c3mdci} as well as in \eqref{triplecut}, the loop legs are evaluated on the solution of the on-shell conditions for the cut legs. Since the three-particle cut is not maximal in four dimensions, the on-shell constraints fix a one-parameter family of solutions and do not allow to fix immediately the coefficient of the three-mass triangle. Geometrically, this corresponds to a curve of allowed values for the internal region variable $x_0$. This is the curve of points that are light-like separated from the three points $x_a$, $x_b$ and $x_c$. 

The construction of \cite{Forde:2007mi} showed that there is a particular value on this curve that isolates the triangle coefficient. Furthermore, since the constraint is quadratic, there are  two solutions and, as dictated by generalised unitarity, one has to take an  average. Details on this procedure are provided in 
Appendix~\ref{solutiontriangle}. To simplify the final result, it is convenient to introduce the variables
\begin{align}\label{uandv}
\frac{x^2_{ab}}{x^2_{ac}}&= u= z \bar z \;, &   \frac{x^2_{bc}}{x^2_{ac}}&= v = (1-z)(1-\bar z) \;.
\end{align}
In terms of these variables, the coefficient of the triangle can be cast  in the form
\begin{align}\label{c3msol}
c^{\text{3m}}=\frac{1}{\D_{abc}}\left[ \frac{\langle r-1\,r\rangle \langle s-1\,s\rangle \delta^{(4)}((z-1)\langle K^\flat \, \theta_{ab}\rangle + z\langle K^\flat \, \theta_{bc}\rangle)}{z(1-z)\langle r\,K^\flat\rangle \langle s-1\,K^\flat\rangle \langle s\,K^\flat\rangle \langle t\,K^\flat\rangle}  + (z\leftrightarrow\bar{z}) \right]\, , 
\end{align}
with%
\footnote{Compared to \cite{Bork:2012tt}, our definition of $K^{\flat}$ is rescaled for convenience, taking advantage of cancellations between numerator and denominator (see also \eqref{KflatK1flat}).}
\begin{align}\label{Kflat}
 K^{\flat \mu}=x_{ab}^{\mu} (z-1)+ x_{bc}^{\mu} z.
\end{align}
Notice that $(K^{\flat})^2=0$, which allows us to use it inside the spinor brackets. The sum over the exchange of $z$ and $\bar z$ in  \eqref{c3msol} corresponds to the average over the two solutions  discussed earlier and it involves also the definition of $K^{\flat}$.

The exchange of $z$ and $\bar z$ is not the only symmetry of $c^{\text{3m}}$. It is  easy to see that \eqref{c3msol} is symmetric under the exchange 
\begin{align}\label{exchange}
\left\{
\begin{aligned}
x_{ab} &\leftrightarrow x_{bc}\ ,  \cr
u &\leftrightarrow v\ .
\end{aligned}\right.
\end{align}
This particular feature will be important in the following.

The form \eqref{c3msol} is not ideal to test dual conformal invariance. We will find an alternative expression which makes  this symmetry more manifest. In order to achieve this, we start from \eqref{c3mdci}. Importantly, we will not need the particular form of the solution to prove dual conformal symmetry. In other words, our derivation applies for any $x_0$ sitting on the curve of solutions to the on-shell conditions for the three cut legs. As a bonus, we will see that this  derivation allows an easier evaluation on the kinematic solution with respect to \eqref{c3mdci}. First we rewrite \eqref{c3mdci} using the identities
\begin{align}
 \braket{\ell_2 \, \ell_1}[\ell_1 \, \ell_3]\braket{\ell_3 \, r-1}&=\braket{\ell_2|x_{0a}x_{ac}|r-1}\, ,  & \braket{\ell_2 \, \ell_3}[\ell_3 \, \ell_1]\braket{\ell_1 \, r}&=\braket{\ell_2|x_{0c}x_{ca}|r} \, ,\\
  \braket{\ell_2 \, \ell_1}[\ell_1 \, \ell_2]\braket{\ell_2 \, s-1}&=\braket{\ell_2|x_{0a}x_{ab}|s-1} \, ,& \braket{\ell_2 \, \ell_3}[\ell_3 \, \ell_2]\braket{\ell_2 \, s}&=\braket{\ell_2|x_{0c}x_{cb}|s} \, ,\\
  \braket{ \ell_2 \, \ell_1}[\ell_1 \, \ell_3]\braket{\ell_3 \, \theta_{bc}}&=-\braket{\ell_2 |x_{0a}x_{ac} |\theta_{cb}} \, ,& \braket{ \ell_2 \, \ell_3}[\ell_3 \, \ell_1]\braket{\ell_1 \, \theta_{ab}}&=\braket{\ell_2 |x_{0c}x_{ca} | \theta_{ab}}\, ,
\end{align}
where we used momentum conservation at the three vertices and the on-shell condition for the loop legs. Furthermore, the loop leg $\ell_2$ is adjacent both to $x_0$ and $x_b$, therefore
\begin{align}
 \bra{\ell_2} x_0=\bra{\ell_2} x_b \ .
\end{align}
This gives
\begin{align}\label{c3mintermediate}
 c^{\text{3m}}=\frac{\langle s-1\,s\rangle \langle r-1\,r\rangle \,\d^{(4)}(\braket{ \ell_2 |x_{ba}x_{ac} |\theta_{cb}}+ \braket{ \ell_2 |x_{bc}x_{ca} | \theta_{ab}})}{x_{ac}^2 \braket{\ell_2|x_{ba}x_{ab}|s-1} \braket{\ell_2|x_{bc}x_{cb}|s}\braket{\ell_2|x_{ba}x_{ac}|r-1}\braket{\ell_2|x_{bc}x_{ca}|r}} \frac{uv}{\Delta}\, , 
\end{align}
where we introduced the quantity 
\begin{align}
 \Delta = \sqrt{(1-u-v)^2-4uv} =|z-\bar z|
 \ .
\end{align}
Using momentum supertwistors and the identities 
\begin{align}
 \braket{s-1\, s} x^2_{ab} \braket{r-1\, r}&= -\braket{s-1,s,r-1,r}\, ,  
 \\ 
 \braket{s-1\, s} x^2_{bc} \braket{r-1\, r}&= -\braket{s-1,s,(r-1)^-,r^-}\, ,  \\ 
 \braket{r-1\, r}^2 x^2_{ac}&=-\braket{r-1,r,(r-1)^-,r^-}\ , 
\end{align}
we can rewrite \eqref{c3mintermediate} as
\begin{align}
\label{445}
 c^{\text{3m}} = \mathcal{R}_{r,s}(\ell_2) \frac{\sqrt{u v}}{\D}\ , 
\end{align}
with 
\begin{align}\label{Rrs}
 \mathcal{R}_{r,s}(\ell_2)=\;&[\ell_2,r,r-1,r^-,(r-1)^-] \; \frac{\braket{\ell_2,r,r-1,r^-}\braket{\ell_2,r^-,(r-1)^-,r-1}}{\braket{\ell_2,r,r-1,s-1}\braket{\ell_2,r^-,(r-1)^-,s}} \cr
 & \; \times\frac{\braket{s-1,s,r-1,r}^{\frac12} \braket{s-1,s,(r-1)^-,r^-}^{\frac12}}{\braket{r-1,r,(r-1)^-1,r^-}}
 \ .
\end{align}
To arrive at this expression in terms of dual conformal invariant five- and four-brackets, we introduced the new supertwistor
\begin{align}
 \mathcal{Z}^M_{\ell_2}&=\begin{pmatrix} Z^{\hat A}_{\ell_2} \\ \theta_b^{A\alpha} \lambda_{\ell_2 \alpha} \end{pmatrix} \, , & Z^{\hat A}_{\ell_2}=\begin{pmatrix} \lambda_{\ell_2}^{\alpha} \\ x_b^{\dot \alpha \alpha} \lambda_{\ell_2 \alpha} \end{pmatrix}\ .
\end{align}
One can easily check that \eqref{Rrs} is invariant under the little group scaling \eqref{scaling} as well.

 The emergence of dual conformal invariant structures in a three-particle cut is a pleasant surprise and a strong hint of dual conformal invariance. As we already stressed, \eqref{445}  is to be evaluated at a specific value of the loop momenta. Notice, however, that in this version of $c^{\text{3m}}$ the whole dependence on the loop momenta is through $\lambda_{\ell_2}$. Therefore it is extremely simple to evaluate it on the explicit solution. Indeed, as we review in Appendix~\ref{solutiontriangle}, in the limit corresponding to the direct extraction of the triangle coefficient one can effectively replace
 \begin{align}\label{l2Kflat}
  \lambda_{\ell_2}\ \to \  \lambda_{K^{\flat}}\, , 
 \end{align}
with $K^{\flat}$ given in \eqref{Kflat}. With this insight, we can finally write
\begin{align}\label{c3mnew}
  c^{\text{3m}} =\frac12\left(\mathcal{R}_{r,s}(K^{\flat})+\mathcal{R}_{r,s}(\bar{K}^{\flat})\right) \frac{\sqrt{u v}}{\D} 
\ , 
\end{align}
where $\bar K^{\flat}$ is obtained from $K^{\flat}$ after the replacement $z\to \bar z$. $K^{\flat}$ and $\bar K^{\flat}$ correspond to the two solutions of the on-shell constraints. Although it is not immediately obvious,  \eqref{c3mnew} and \eqref{c3msol} are identical.

 After fixing this coefficient,  we are left with
\begin{align}
c^{\text{3m}} \; \raisebox{-.47\height}{
\begin{tikzpicture}[thick, scale=0.7]
\coordinate (T1) at ( 0.5 ,  0.87);
\coordinate (T2) at ( 0.5 , -0.87);
\coordinate (T3) at (-1.0 ,  0.0 );
\draw (T1) -- (T2) -- (T3) -- (T1);
\node[] (Tlbl3) at (-0.6 ,  0.8) {$x_a$};
\node[] (Tlbl1) at ( 1.2 ,  0.0) {$x_b$};
\node[] (Tlbl2) at (-0.6 , -0.8) {$x_a^-$};
\node[] (Tlbl0) at ( 0.0 ,  0.0) {};
\draw (T1) -- ++( 60:0.8) node[above right=-2pt] {$r$};
\draw[dotted] (T1)+( 50:0.6) to [bend left=45] ++( 10:0.6);
\draw (T1) -- ++(  0:0.8) node[right] {$s-1$};
\draw (T2) -- ++(  0:0.8) node[right] {$s$};
\draw[dotted] (T2)+(-10:0.6) to [bend left=45] ++(-50:0.6);
\draw (T2) -- ++(-60:0.8) node[below right=-2pt] {$r-1$};
\draw[double] (T3) -- ++(180:0.8);
\end{tikzpicture}
}&= \; \frac12\left(\mathcal{R}_{r,s}(K^{\flat})+\mathcal{R}_{r,s}(\bar{K}^{\flat})\right) g(u,v)\, , 
\end{align}
where
\begin{align}\label{gdef}
 g(u,v)=\frac{\sqrt{uv}}{\Delta} F^{3\mathrm{m}} (z,\bar z)\, ,  
\end{align}
and $F^{3\mathrm{m}}(z,\bar z)$ is the explicit result of the reduced three-mass triangle (see Appendix \ref{integrals}) 
\begin{align}\label{F3m}
F^{3\mathrm{m}}(z,\bar z) = \Li(z)-\Li(\bar z)+\tfrac{1}{2}\log(z \bar z)\log\left(\frac{1-z}{1-\bar z}\right) \;.
\end{align}

What remains to be proven is the invariance of the function $g(u,v)$. However it is not hard to see, by acting with the generator $\mathsf{K}^\mu$ in \eqref{Kmu}, that the variation of $g(u,v)$ is non-vanishing. On the other hand, we will now show that this variation cancels in the sum over all possible triangles. 
To begin with, one can show that $F^{3\mathrm{m}}(z,\bar z)=F^{3\mathrm{m}}(1-z,1-\bar z)$
as a consequence of the identity 
\begin{align}
 \text{Li}_2(z)=-\text{Li}_2(1-z)-\log
   (1-z) \log (z)+\frac{\pi ^2}{6}\, , 
\end{align}
 thus implying
\begin{align}\label{symmetry}
 g(u,v)=g(v,u)\, .
\end{align}
Therefore $g(u,v)$ is a symmetric function under the exchange \eqref{exchange}. Notice that, in the sum over all possible three-mass triangles one always has a contribution where $u$ and $v$ are swapped. These are 
\begin{align}
c^{\text{3m}} \; \raisebox{-.47\height}{
\begin{tikzpicture}[thick, scale=0.6]
\coordinate (T1) at ( 0.5 ,  0.87);
\coordinate (T2) at ( 0.5 , -0.87);
\coordinate (T3) at (-1.0 ,  0.0 );
\draw (T1) -- (T2) -- (T3) -- (T1);
\node[] (Tlbl3) at (-0.6 ,  0.8) {$x_a$};
\node[] (Tlbl1) at ( 1.2 ,  0.0) {$x_b$};
\node[] (Tlbl2) at (-0.6 , -0.8) {$x_a^-$};
\node[] (Tlbl0) at ( 0.0 ,  0.0) {};
\draw (T1) -- ++( 60:0.8) node[above right=-2pt] {$r$};
\draw[dotted] (T1)+( 50:0.6) to [bend left=45] ++( 10:0.6);
\draw (T1) -- ++(  0:0.8) node[right] {$s-1$};
\draw (T2) -- ++(  0:0.8) node[right] {$s$};
\draw[dotted] (T2)+(-10:0.6) to [bend left=45] ++(-50:0.6);
\draw (T2) -- ++(-60:0.8) node[below right=-2pt] {$r-1$};
\draw[double] (T3) -- ++(180:0.8);
\end{tikzpicture}
}&= \; \frac12\left(\mathcal{R}_{r,s}(K^{\flat})+\mathcal{R}_{r,s}(\bar{K}^{\flat})\right) g(u,v) \\
c^{\text{3m}} \; \raisebox{-.47\height}{
\begin{tikzpicture}[thick, scale=0.6]
\coordinate (T1) at ( 0.5 ,  0.87);
\coordinate (T2) at ( 0.5 , -0.87);
\coordinate (T3) at (-1.0 ,  0.0 );
\draw (T1) -- (T2) -- (T3) -- (T1);
\node[] (Tlbl3) at (-0.6 ,  0.8) {$x_b$};
\node[] (Tlbl1) at ( 1.2 ,  0.0) {$x_a^-$};
\node[] (Tlbl2) at (-0.6 , -0.8) {$x_b^-$};
\node[] (Tlbl0) at ( 0.0 ,  0.0) {};
\draw (T1) -- ++( 60:0.8) node[above right=-2pt] {$s$};
\draw[dotted] (T1)+( 50:0.6) to [bend left=45] ++( 10:0.6);
\draw (T1) -- ++(  0:0.8) node[right] {$r-1$};
\draw (T2) -- ++(  0:0.8) node[right] {$r$};
\draw[dotted] (T2)+(-10:0.6) to [bend left=45] ++(-50:0.6);
\draw (T2) -- ++(-60:0.8) node[below right=-2pt] {$s-1$};
\draw[double] (T3) -- ++(180:0.8);
\end{tikzpicture}
}&= \; \frac12\left(\mathcal{R}_{r,s}(K^{\flat})+\mathcal{R}_{r,s}(\bar{K}^{\flat})\right) g(v,u)
\end{align}
where we used the property $\mathcal{R}_{r,s}=\mathcal{R}_{s,r}$, which we mentioned around \eqref{exchange}.
Crucially  these two configurations are identical when written in terms of Mandelstam invariants, but it is immediate to see that their region variables assignments are different and consequently also their variation under dual special conformal transformation. In particular, we will show that 
\begin{align}\label{antisymmetry}
 \mathsf{K}^\mu g(u,v)=-\mathsf{K}^\mu g(v,u) \, , 
\end{align}
thus providing the cancellation
\begin{align}\label{cancellation}
\mathsf{K}^\mu\left(c^{\text{3m}} \; \raisebox{-.47\height}{
\begin{tikzpicture}[thick, scale=0.5]
\coordinate (T1) at ( 0.5 ,  0.87);
\coordinate (T2) at ( 0.5 , -0.87);
\coordinate (T3) at (-1.0 ,  0.0 );
\draw (T1) -- (T2) -- (T3) -- (T1);
\node[] (Tlbl3) at (-0.6 ,  0.8) {$x_a$};
\node[] (Tlbl1) at ( 1.2 ,  0.0) {$x_b$};
\node[] (Tlbl2) at (-0.6 , -0.8) {$x_a^-$};
\node[] (Tlbl0) at ( 0.0 ,  0.0) {};
\draw (T1) -- ++( 60:0.8) node[above right=-2pt] {$r$};
\draw[dotted] (T1)+( 50:0.6) to [bend left=45] ++( 10:0.6);
\draw (T1) -- ++(  0:0.8) node[right] {$s-1$};
\draw (T2) -- ++(  0:0.8) node[right] {$s$};
\draw[dotted] (T2)+(-10:0.6) to [bend left=45] ++(-50:0.6);
\draw (T2) -- ++(-60:0.8) node[below right=-2pt] {$r-1$};
\draw[double] (T3) -- ++(180:0.8);
\end{tikzpicture}
}+ \; c^{\text{3m}} \; \raisebox{-.47\height}{
\begin{tikzpicture}[thick, scale=0.5]
\coordinate (T1) at ( 0.5 ,  0.87);
\coordinate (T2) at ( 0.5 , -0.87);
\coordinate (T3) at (-1.0 ,  0.0 );
\draw (T1) -- (T2) -- (T3) -- (T1);
\node[] (Tlbl3) at (-0.6 ,  0.8) {$x_b$};
\node[] (Tlbl1) at ( 1.2 ,  0.0) {$x_a^-$};
\node[] (Tlbl2) at (-0.6 , -0.8) {$x_b^-$};
\node[] (Tlbl0) at ( 0.0 ,  0.0) {};
\draw (T1) -- ++( 60:0.8) node[above right=-2pt] {$s$};
\draw[dotted] (T1)+( 50:0.6) to [bend left=45] ++( 10:0.6);
\draw (T1) -- ++(  0:0.8) node[right] {$r-1$};
\draw (T2) -- ++(  0:0.8) node[right] {$r$};
\draw[dotted] (T2)+(-10:0.6) to [bend left=45] ++(-50:0.6);
\draw (T2) -- ++(-60:0.8) node[below right=-2pt] {$s-1$};
\draw[double] (T3) -- ++(180:0.8);
\end{tikzpicture}
}\right)=0\, .
\end{align}
In order to prove our crucial result \eqref{antisymmetry}, we start from the variation of the basic ingredients
\begin{align}\label{uvvar}
\mathsf{K}^\mu u = -2u \, x_{bc}^\mu \;, \qquad \mathsf{K}^\mu v = -2v \, x_{ba}^\mu \;,
\end{align}
from which we derive
\begin{align}
 \mathsf{K}^\mu g(u,v)=-2 u\partial_u g(u,v) x_{bc}^\mu +2 v\partial_v g(u,v) x_{ab}^\mu \, .
\end{align}
Now we apply to this equation the exchange \eqref{exchange}, leading to
\begin{align}
 \mathsf{K}^\mu g(v,u)=-2 v\partial_v g(v,u) x_{ab}^\mu +2 u\partial_u g(v,u) x_{bc}^\mu \, .
\end{align}
Then, we can simply use the identities
\begin{align}
 \partial_u g(v,u)&=\partial_u g(u,v) \, , & \partial_v g(v,u)&=\partial_v g(u,v)\, , 
\end{align}
which are trivial consequences of \eqref{symmetry},  to see that \eqref{antisymmetry} holds for any symmetric function of $u$ and $v$. 

In summary, we have proven that, given a symmetric function of $u$ and $v$, its variation under dual conformal transformation is antisymmetric in $u$ and $v$. In particular this applies to $g(u,v)$ defined in \eqref{gdef} (for completeness we have written its explicit variation in Appendix~\ref{variations}). Therefore, we conclude that the variation of the three-mass triangle contributions cancels out in the sum over all the possible triangles. 
We stress how non-trivial this result is -- quantities involving triangle functions can therefore be dual conformal invariant. 

As an example, let us discuss in detail the four-point case.  In that case one simply has four possible permutations,  and the cancellation is

\begin{align}
\mathsf{K}^\mu\left(c^{\text{3m}} \; \raisebox{-.47\height}{
\begin{tikzpicture}[thick, scale=0.5]
\coordinate (T1) at ( 0.5 ,  0.87);
\coordinate (T2) at ( 0.5 , -0.87);
\coordinate (T3) at (-1.0 ,  0.0 );
\draw (T1) -- (T2) -- (T3) -- (T1);
\node[] (Tlbl3) at (-0.6 ,  0.8) {$x_1$};
\node[] (Tlbl1) at ( 1.2 ,  0.0) {$x_3$};
\node[] (Tlbl2) at (-0.6 , -0.8) {$x_1^-$};
\node[] (Tlbl0) at ( 0.0 ,  0.0) {};
\draw (T1) -- ++( 60:0.8) node[above right=-2pt] {$1$};
\draw (T1) -- ++(  0:0.8) node[right] {$2$};
\draw (T2) -- ++(  0:0.8) node[right] {$3$};
\draw (T2) -- ++(-60:0.8) node[below right=-2pt] {$4$};
\draw[double] (T3) -- ++(180:0.8);
\end{tikzpicture}
}+ \; c^{\text{3m}} \; \raisebox{-.47\height}{
\begin{tikzpicture}[thick, scale=0.5]
\coordinate (T1) at ( 0.5 ,  0.87);
\coordinate (T2) at ( 0.5 , -0.87);
\coordinate (T3) at (-1.0 ,  0.0 );
\draw (T1) -- (T2) -- (T3) -- (T1);
\node[] (Tlbl3) at (-0.6 ,  0.8) {$x_3$};
\node[] (Tlbl1) at ( 1.2 ,  0.0) {$x_1^-$};
\node[] (Tlbl2) at (-0.6 , -0.8) {$x_3^-$};
\node[] (Tlbl0) at ( 0.0 ,  0.0) {};
\draw (T1) -- ++( 60:0.8) node[above right=-2pt] {$3$};
\draw (T1) -- ++(  0:0.8) node[right] {$4$};
\draw (T2) -- ++(  0:0.8) node[right] {$1$};
\draw (T2) -- ++(-60:0.8) node[below right=-2pt] {$2$};
\draw[double] (T3) -- ++(180:0.8);
\end{tikzpicture}
}\right)&=0 \\
\mathsf{K}^\mu\left(c^{\text{3m}} \; \raisebox{-.47\height}{
\begin{tikzpicture}[thick, scale=0.5]
\coordinate (T1) at ( 0.5 ,  0.87);
\coordinate (T2) at ( 0.5 , -0.87);
\coordinate (T3) at (-1.0 ,  0.0 );
\draw (T1) -- (T2) -- (T3) -- (T1);
\node[] (Tlbl3) at (-0.6 ,  0.8) {$x_2$};
\node[] (Tlbl1) at ( 1.2 ,  0.0) {$x_4$};
\node[] (Tlbl2) at (-0.6 , -0.8) {$x_2^-$};
\node[] (Tlbl0) at ( 0.0 ,  0.0) {};
\draw (T1) -- ++( 60:0.8) node[above right=-2pt] {$2$};
\draw (T1) -- ++(  0:0.8) node[right] {$3$};
\draw (T2) -- ++(  0:0.8) node[right] {$4$};
\draw (T2) -- ++(-60:0.8) node[below right=-2pt] {$1$};
\draw[double] (T3) -- ++(180:0.8);
\end{tikzpicture}
} + \; c^{\text{3m}} \; \raisebox{-.47\height}{
\begin{tikzpicture}[thick, scale=0.5]
\coordinate (T1) at ( 0.5 ,  0.87);
\coordinate (T2) at ( 0.5 , -0.87);
\coordinate (T3) at (-1.0 ,  0.0 );
\draw (T1) -- (T2) -- (T3) -- (T1);
\node[] (Tlbl3) at (-0.6 ,  0.8) {$x_4$};
\node[] (Tlbl1) at ( 1.2 ,  0.0) {$x_2^-$};
\node[] (Tlbl2) at (-0.6 , -0.8) {$x_4^-$};
\node[] (Tlbl0) at ( 0.0 ,  0.0) {};
\draw (T1) -- ++( 60:0.8) node[above right=-2pt] {$4$};
\draw (T1) -- ++(  0:0.8) node[right] {$1$};
\draw (T2) -- ++(  0:0.8) node[right] {$2$};
\draw (T2) -- ++(-60:0.8) node[below right=-2pt] {$3$};
\draw[double] (T3) -- ++(180:0.8);
\end{tikzpicture}
}\right)&=0
\end{align}
which can be checked explicitly. 

\section{Conclusions and outlook}

In this paper we provided strong evidence for the invariance of quantum form factors under dual conformal symmetry. At tree level, this was partly understood in  \cite{Bork:2014eqa} using a formulation in terms of twistor variables. The extension of these results to loop level seemed to be obstructed by the appearance of scalar triangles in the loop expansion. Here, we presented a general argument for one-loop dual conformal invariance and  explicitly analysed the cancellation mechanism leading to a vanishing variation for finite triangles. Importantly, our results  rely on the  prescription  introduced in \cite{Part1} to express the integrated result in terms of dual region momenta.

Our observation opens the way to many future developments. One obvious question is whether dual conformal invariance survives for higher loops and, if so, which constraints can be put on the allowed scalar integrals and their coefficients. At one loop we already noticed interesting features. In \eqref{NMHV4pt1loop} the box integrals organise themselves in a simple combination, whose variation under dual conformal symmetry yields exactly the correct anomaly \eqref{KV}. Conversely, one could say that dual conformal invariance constrains the box coefficients such that the combination of box functions leads to the correct dual conformal anomaly. 
 A similar argument allows to exclude the presence of three-mass triangles different from \eqref{3mscalartriangle}. Indeed, while cancellations like \eqref{cancellation} do not rely on having only the off-shell leg in one corner, the possibility of recasting the three-mass coefficient in a dual conformal invariant form (such as  \eqref{Rrs}), is linked to the specific configuration \eqref{3mscalartriangle} where the the off-shell leg sits alone at one corner, and would be spoiled in a more general case.

Another interesting question is whether dual conformal invariance survives for form factors of different operators. One could start looking at protected longer operators, for which some loop results are already available \cite{Brandhuber:2014ica,Penante:2014sza}. Afterwards, one would naturally move to unprotected operators \cite{Brandhuber:2016fni,Loebbert:2016xkw,Chicherin:2016qsf,Brandhuber:2017bkg,Brandhuber:2018xzk,Brandhuber:2018kqb}. In that case the presence of ultraviolet divergences makes things more subtle  and the argument of Section \ref{oneloop} would have to be revisited.

Since our method for showing dual conformal invariance applies to the expansion of the result in terms of scalar integrals, it would be  important to develop a general method to test dual conformal symmetry on the final result in terms of Mandelstam invariants. In particular, while there is an unambiguous map between Mandelstam invariants and region variables, the definition of $q^2$ (and in particular its variation under dual conformal invariance) changes according to the specific scalar integral. Rewriting Mandelstam variables in terms of twistors may potentially help in finding new dual conformal invariants on the periodic configuration.

It would also be exciting   to understand the precise Wilson loop dual of form factors. In the dual picture, dual conformal invariance is simply the ordinary conformal invariance of the Wilson loop expectation value and this would provide new important insights. In particular, given the latest developments in the computation of exact scattering amplitudes, a Wilson loop dual would allow to access the non-perturbative regime, thus gaining a deeper understanding of the symmetries. 

We conclude by mentioning one last future direction that we would like to investigate. As we mentioned in the Introduction, the authors of \cite{Bourjaily:2013mma} developed a dual conformal invariant regularisation for the case of scattering amplitudes. This led to the formulation of new unitarity-based techniques which allow to compute the integrand of scattering amplitudes for arbitrary helicity configurations and number of external legs up to three loops \cite{Bourjaily:2017wjl}. A similar technique for the case of form factors would allow to notably increase the amount of perturbative data at our disposal.

\section*{Acknowledgements}

We would like to thank \"{O}mer Gurdo\u{g}an, Florian Loebbert, Jan Plefka and Alexander Tumanov for interesting discussions. 
The work of LB is supported by a Marie Sk\l{}odowska-Curie Individual Fellowship under grant agreement No.~749909.
The work of AB and GT was supported by the Science and Technology Facilities Council (STFC) Consolidated Grant ST/P000754/1  
\textit{``String theory, gauge theory \& duality"}. This project has received funding from the European Union's Horizon 2020 research and innovation programme under the Marie Sk\l{}odowska-Curie grant agreement 
No.~764850 {\it ``SAGEX"}.
GT is grateful to the Alexander von Humboldt Foundation for support through a Friedrich Wilhelm Bessel Research Award, and to the Institute for Physics and IRIS Adlershof at Humboldt University, Berlin, for their warm hospitality.
LB would like to thank the Dipartimento di Fisica of Torino University for kind hospitality during the initial phase of this project.
	
\appendix
\section{Notation and conventions} \label{conventions}
Throughout this paper we use the following notation to indicate the N$^k$MHV tree-level amplitude and form factor
\begin{align}
F^{(0)}_{n,k} &=
\raisebox{-.45\height}{
\begin{tikzpicture}[thick]
  \greyblob (blob) at (0,0) {$\scriptstyle{k}$};
  \draw[double] (blob) -- ++( 180:0.8) node[left] {};
  \draw (blob) -- ++( 135:0.8) node[above left=-2pt] {$1$};
  \draw (blob) -- ++(  45:0.8) node[above right=-2pt] {$2$};
  \draw[dotted] (blob)+(  35:0.6) to [bend left=45] ++( -35:0.6);
  \draw (blob) -- ++( -45:0.8) node[below right=-4pt] {$n-1$};
  \draw (blob) -- ++(-135:0.8) node[below left=-2pt] {$n$};
\end{tikzpicture}
} &
A^{(0)}_{n,k} &=
\raisebox{-.45\height}{
\begin{tikzpicture}[thick]
  \greyblob (blob) at (0,0) {$\scriptstyle{k}$};
  \draw (blob) -- ++( 135:0.8) node[above left=-2pt] {$1$};
  \draw (blob) -- ++(  45:0.8) node[above right=-2pt] {$2$};
  \draw[dotted] (blob)+(  35:0.6) to [bend left=45] ++( -35:0.6);
  \draw (blob) -- ++( -45:0.8) node[below right=-4pt] {$n-1$};
  \draw (blob) -- ++(-135:0.8) node[below left=-2pt] {$n$};
\end{tikzpicture}
}
\end{align}
Our conventions for the MHV cases are
\begin{align}
F^{(0)}_{n,0} & = \frac{\delta^{(8)}(\gamma-{\sum_{i=1}^n \lambda_i \eta_i})}{\braket{1\,2}\cdots \braket{n\,1}} \;,
&
A^{(0)}_{n,0} & = \mathrm{i}\frac{ \delta^{(8)}({\sum_{i=1}^n \lambda_i \eta_i })}{\braket{1\,2}\cdots \braket{n\,1}}
\end{align}
The usual delta function for momentum conservation is not indicated. For the simplest cases of three-point amplitude and two-point form factor we use the notation
\begin{align}
A_{3,0}^{(0)} &=
\raisebox{-.44\height}{
\begin{tikzpicture}[thick]
  \blackdot (MHVb) at (0.0, 0.0) {};
  \draw (MHVb) -- ++( 120:0.8) node[anchor=east] {$1$};
  \draw (MHVb) -- ++(   0:0.8) node[anchor=west] {$2$};
  \draw (MHVb) -- ++(-120:0.8) node[anchor=east] {$3$};
\end{tikzpicture}
} = \mathrm{i} \, \frac{\delta^{(8)}(\mathrm{\lambda_1\eta_1  + \lambda_2\eta_2 +\lambda_3\eta_3
})}{\langle1\,2\rangle \langle2\,3\rangle \langle3\,1\rangle}\, ,  \cr
A_{3,-1}^{(0)} &=
\raisebox{-.44\height}{
\begin{tikzpicture}[thick]
  \whitedot (MHVb) at (0.0, 0.0) {};
  \draw (MHVb) -- ++( 120:0.8) node[anchor=east] {$1$};
  \draw (MHVb) -- ++(   0:0.8) node[anchor=west] {$2$};
  \draw (MHVb) -- ++(-120:0.8) node[anchor=east] {$3$};
\end{tikzpicture}
} = -\mathrm{i} \, \frac{\delta^{(4)}([2 \, 3]\eta_1 + [3 \, 1]\eta_2 + [1 \, 2]\eta_3)}{[1\,2][2\,3][3\,1]}\, ,  \cr
F_{2,0}^{(0)} &=
\raisebox{-.44\height}{
\begin{tikzpicture}[thick]
  \coordinate (FF) at (0.0, 0.0);
  \draw (FF) -- ++(  60:0.8) node[anchor=west] {$1$};
  \draw (FF) -- ++( -60:0.8) node[anchor=west] {$2$};
  \draw[double] (FF) -- ++( 180:0.8) node[anchor=east] {};
\end{tikzpicture}
} = \frac{\delta^{(8)}(\gamma - \lambda_1\eta_1  - \lambda_2\eta_2)}{\langle1\,2\rangle \langle2\,1\rangle}\, .
\end{align}
All the external legs are outgoing, except for  the off-shell leg. The latter has incoming momentum $q$ and supermomentum $\gamma$, with 
\begin{align}
q = \sum_{i=1}^n p_i \;, \qquad \gamma = \sum_{i=1}^n \mathrm{q}_i \;.
\end{align}
We use supersymmetric region variables according to the convention
\begin{align}
x^{\alpha\dot{\alpha}}_i - x^{\alpha\dot{\alpha}}_{i+1} &= p_i^{\alpha\dot{\alpha}} = \lambda_i^\alpha\widetilde{\lambda}_i^{\dot{\alpha}} \;, &  \theta^{A \alpha}_i - \theta^{A \alpha}_{i+1} &= \mathrm{q}_i^{A \alpha} = \eta_i^A \lambda_i^{\alpha} \; .
\end{align}
If $q\neq0$ the dual coordinates do not describe a closed polygon. However they are still arranged in periodic configurations, where the image variables are defined as 
\begin{align}\label{periodicity}
x^{[m]}_{i} &= x_i + mq \;,  &  \theta^{[m]}_{i} &= \theta_i + m\gamma \; ,
\end{align}
with $m\in\mathbb{Z}$. For $m=\pm 1$ we use the notation
\begin{align}
 x^\pm_i&=x_i\pm q \, , &  \theta^{\pm}_{i} &= \theta_i \pm \gamma  \; .
 \end{align}
The same kinematic configuration can be encoded in terms of momentum-twistor variables since edges of the periodic line are light rays in dual space. The incidence relation
\begin{align}
\mu^{\dot{\alpha}}_i = x^{\alpha\dot{\alpha}}_i \lambda_{i\,\alpha} = x^{\alpha\dot{\alpha}}_{i+1} \lambda_{i\,\alpha}
\ , 
\end{align}
fixes the components of the twistor 
\begin{align}
Z^{\hat A}_i = \begin{pmatrix} \lambda^{\alpha}_i \\ \mu^{\dot \alpha}_i \end{pmatrix} 
\ , 
\end{align}
and the ambiguity in the choice of the spinor-helicity variables $(\lambda_i, \widetilde{\lambda}_i)$ now translates to the fact that $Z_i$ are interpreted as  projective coordinates in twistor space $\mathbb{T}\simeq\mathbb{CP}^3$. The supersymmetric version is simply 
\begin{align}
 \mathcal{Z}_i^M&=\begin{pmatrix} Z_i^{\hat A}\\ \chi_i^{A}\end{pmatrix} \ , &  \chi^A_i=\theta_i^{A\, \alpha} \lambda_{i \alpha} \ .
\end{align}
Periodicity, as in \eqref{periodicity}, is implemented by the condition
\begin{align}
 \mathcal{Z}^{[m] M}_i&=\begin{pmatrix} Z_i^{[m] \hat A}\\ \chi_i^{[m] A} \end{pmatrix} \ , & Z_i^{[m]\hat A}&=\begin{pmatrix} \lambda_i^{\alpha}\\ (x_i^{[m]})^{\dot \alpha \alpha} \lambda_{i\alpha} \end{pmatrix} \ , & \chi_i^{[m] A}=(\theta_i^{[m]})^{A\alpha} \lambda_{i \alpha} \ . 
\end{align}
In Section \ref{tree_level} we introduced the following notation for $R$-invariants:
\begin{align}R_{rst}=\raisebox{-.45\height}{
\begin{tikzpicture}[thick, scale=0.8]
  \drawLLwhite
  \drawUL{$\scriptstyle{0}$}
  \drawUR{$\scriptstyle{0}$}
  \drawLR{$\scriptstyle{0}$}
  \drawregionvariables{$x_c$}{$x_{c+1}$}{$x_a$}{$x_b$}{}
  \drawboxinternallines
  \draw (UL) -- ++( 180:0.8) node[anchor=east] {$r+1$};
  \draw[dotted] (UL)+( 170:0.6) to [bend left=45] ++( 100:0.6);
  \draw (UL) -- ++(  90:0.8) node[anchor=south] {$s-1$};
  \draw (UR) -- ++( 0:0.8) node[anchor= west] {$t-1$};
  \draw[dotted] (UR)+(  90:0.6) to [bend left=45] ++(  0:0.6);
  \draw (UR) -- ++(  90:0.8) node[anchor=south] {$s$};
  \draw (LR) -- ++(   0:0.8) node[anchor=west] {$t$};
  \draw[dotted] (LR)+( -10:0.6) to [bend left=45] ++( -80:0.6);
  \draw (LR) -- ++( -90:0.8) node[anchor=north] {$r-1$}; 
  \draw (LL) -- ++(-135:0.8) node[anchor=north east] {$r$};
\end{tikzpicture}}
\end{align}
hinting at their connection  to a quadruple cut. The precise relation is the following
\begin{align}\label{cuttoR}
\raisebox{-.45\height}{
\begin{tikzpicture}[thick, scale=0.8]
  \drawLLwhite
  \drawUL{$\scriptstyle{0}$}
  \drawUR{$\scriptstyle{0}$}
  \drawLR{$\scriptstyle{0}$}
  \drawregionvariables{$x_c$}{$x_{c+1}$}{$x_a$}{$x_b$}{}
  \draw (UL) --node[circle, fill=white, draw=white, inner sep=2pt]{} (UR) --node[circle, fill=white, draw=white, inner sep=2pt]{} (LR) -- (LL) -- (UL);
  \node[circle, fill=white, draw=white, inner sep=2pt] at (-\boxsize,0) {};
  \node[circle, fill=white, draw=white, inner sep=2pt] at (0,-\boxsize) {};
  \draw[red, dashed] (0,0)+(  0:0.5) -- ++(  0:0.9)  (0,0)+( 90:0.5) -- ++( 90:0.9)  (0,0)+(180:0.5) -- ++(180:0.9)  (0,0)+(-90:0.5) -- ++(-90:0.9);
  \draw (UL) -- ++( 180:0.8) node[anchor=east] {$r+1$};
  \draw[dotted] (UL)+( 170:0.6) to [bend left=45] ++( 100:0.6);
  \draw (UL) -- ++(  90:0.8) node[anchor=south] {$s-1$};
  \draw (UR) -- ++( 0:0.8) node[anchor= west] {$t-1$};
  \draw[dotted] (UR)+(  90:0.6) to [bend left=45] ++(  0:0.6);
  \draw (UR) -- ++(  90:0.8) node[anchor=south] {$s$};
  \draw (LR) -- ++(   0:0.8) node[anchor=west] {$t$};
  \draw[dotted] (LR)+( -10:0.6) to [bend left=45] ++( -80:0.6);
  \draw (LR) -- ++( -90:0.8) node[anchor=north] {$r-1$}; 
  \draw (LL) -- ++(-135:0.8) node[anchor=north east] {$r$};
\end{tikzpicture}}
= \mathrm{i} \Delta_{a b c\; c+1} A^{(0)}_{n,0} \raisebox{-.45\height}{\begin{tikzpicture}[thick, scale=0.8]
  \drawLLwhite
  \drawUL{$\scriptstyle{0}$}
  \drawUR{$\scriptstyle{0}$}
  \drawLR{$\scriptstyle{0}$}
   \drawregionvariables{$x_c$}{$x_{c+1}$}{$x_a$}{$x_b$}{}
  \drawboxinternallines
  \draw (UL) -- ++( 180:0.8) node[anchor=east] {$r+1$};
  \draw[dotted] (UL)+( 170:0.6) to [bend left=45] ++( 100:0.6);
  \draw (UL) -- ++(  90:0.8) node[anchor=south] {$s-1$};
  \draw (UR) -- ++( 0:0.8) node[anchor= west] {$t-1$};
  \draw[dotted] (UR)+(  90:0.6) to [bend left=45] ++(  0:0.6);
  \draw (UR) -- ++(  90:0.8) node[anchor=south] {$s$};
  \draw (LR) -- ++(   0:0.8) node[anchor=west] {$t$};
  \draw[dotted] (LR)+( -10:0.6) to [bend left=45] ++( -80:0.6);
  \draw (LR) -- ++( -90:0.8) node[anchor=north] {$r-1$}; 
  \draw (LL) -- ++(-135:0.8) node[anchor=north east] {$r$};
\end{tikzpicture}}
\end{align}
with
\begin{align}\label{Delta4m}
 \Delta_{a b c d}= \sqrt{(x_{ac}^2 x_{bd}^2-x_{bc}^2 x_{ad}^2+x_{ab}^2 x_{cd}^2)^2 -4x_{ac}^2 x_{bd}^2 x_{ab}^2 x_{cd}^2 }
 \ .
\end{align}
If $x_{cd}^2=0$, as it happens in \eqref{cuttoR}, this factor reduces to
\begin{align}
  \Delta_{a b c\; c+1}=x_{ac}^2 x_{b c+1}^2 - x_{a c+1}^2 x_{b c}^2 
  \ .
\end{align}
Notice in particular that this is the form of $\Delta_{abcd}$ for all the IR divergent boxes. The four-mass box is the only one for which one needs to use \eqref{Delta4m} and it is IR finite and dual conformal invariant by itself. 

For the case of form factors we have a similar relation between cuts and $R$-invariants
\begin{align}
\raisebox{-.48\height}{
\begin{tikzpicture}[thick, scale=0.8]
  \drawLLwhite
  \drawUL{$\scriptstyle{0}$}
  \drawUR{$\scriptstyle{0}$}
  \drawLR{$\scriptstyle{0}$}
  \drawregionvariables{$x_c$}{$x_{c+1}$}{$x_a$}{$x_b$}{}
  \draw (UL) --node[circle, fill=white, draw=white, inner sep=2pt]{} (UR) --node[circle, fill=white, draw=white, inner sep=2pt]{} (LR) -- (LL) -- (UL);
  \node[circle, fill=white, draw=white, inner sep=2pt] at (-\boxsize,0) {};
  \node[circle, fill=white, draw=white, inner sep=2pt] at (0,-\boxsize) {};
  \draw[red, dashed] (0,0)+(  0:0.5) -- ++(  0:0.9) (0,0)+( 90:0.5) -- ++( 90:0.9) (0,0)+(180:0.5) -- ++(180:0.9) (0,0)+(-90:0.5) -- ++(-90:0.9);
  \draw (UL) -- ++( 180:0.8) node[anchor=east] {$r+1$};
  \draw[dotted] (UL)+( 170:0.6) to [bend left=45] ++( 100:0.6);
  \draw (UL) -- ++(  90:0.8) node[anchor=south] {$s-1$};
  \draw[double] (UR) -- ++(   0:0.8) node[] {};
  \draw (UR) -- ++(  45:0.8) node[anchor=south west] {$t-1$};
  \draw[dotted] (UR)+(  80:0.6) to [bend left=45] ++(  50:0.6);
  \draw (UR) -- ++(  90:0.8) node[anchor=south] {$s$};
  \draw (LR) -- ++(   0:0.8) node[anchor=west] {$t$};
  \draw[dotted] (LR)+( -10:0.6) to [bend left=45] ++( -80:0.6);
  \draw (LR) -- ++( -90:0.8) node[anchor=north] {$r-1$}; 
  \draw (LL) -- ++(-135:0.8) node[anchor=north east] {$r$};
\end{tikzpicture}
}=\mathrm{i} \Delta_{a b c\; c+1} F^{(0)}_{n,0} \raisebox{-.48\height}{
\begin{tikzpicture}[thick, scale=0.8]
  \drawLLwhite
  \drawUL{$\scriptstyle{0}$}
  \drawUR{$\scriptstyle{0}$}
  \drawLR{$\scriptstyle{0}$}
  \drawregionvariables{$x_c$}{$x_{c+1}$}{$x_a$}{$x_b$}{}
  \drawboxinternallines
  \draw (UL) -- ++( 180:0.8) node[anchor=east] {$r+1$};
  \draw[dotted] (UL)+( 170:0.6) to [bend left=45] ++( 100:0.6);
  \draw (UL) -- ++(  90:0.8) node[anchor=south] {$s-1$};
  \draw[double] (UR) -- ++(   0:0.8) node[] {};
  \draw (UR) -- ++(  45:0.8) node[anchor=south west] {$t-1$};
  \draw[dotted] (UR)+(  80:0.6) to [bend left=45] ++(  50:0.6);
  \draw (UR) -- ++(  90:0.8) node[anchor=south] {$s$};
  \draw (LR) -- ++(   0:0.8) node[anchor=west] {$t$};
  \draw[dotted] (LR)+( -10:0.6) to [bend left=45] ++( -80:0.6);
  \draw (LR) -- ++( -90:0.8) node[anchor=north] {$r-1$}; 
  \draw (LL) -- ++(-135:0.8) node[anchor=north east] {$r$};
\end{tikzpicture}
}
\end{align}
and similarly for $R''_{rst}$.

It is well-known that the quadruple cut in four dimensions computes the coefficient of the boxes. The reason why these coefficients in the expansion  \eqref{NMHV4pt1loop} are given in terms of $R$-invariants is that the factor $\mathrm{i} \Delta_{a b c\; c+1}$ is reabsorbed by expanding in a basis of reduced integrals (see Appendix \ref{integrals}), while the tree-level MHV factor cancels when taking the ratio~\eqref{ratio}.

\section{Reduced scalar integrals} \label{integrals}
In this paper we expand one-loop results in terms of reduced scalar integrals, {\it i.e.} conveniently defined dimensionless quantities that are simply related to the original scalar integral. For the boxes we have 
\begin{align}
\frac{1}{2\pi^{2-\eps}r_{\Gamma}}\int \mathrm{d}^{4-2\epsilon} x_0 \; \frac{1}{x_{0a}^2 x_{0b}^2 x_{0c}^2 x_{0d}^2}= \frac{1}{\mathrm i \Delta_{a b c d}} \raisebox{-.47\height}{
\begin{tikzpicture}[thick, scale=0.65]
  \drawLLempty;
  \drawULempty;
  \drawURempty;
  \drawLRempty;
  \drawboxinternallines
  \drawregionvariables{$x_c$}{$x_d$}{$x_a$}{$x_b$}{}
  \draw (UL) -- ++( 150:1.0);
  \draw[dotted] (UL)+( 145:0.8) to [bend left=30] ++( 125:0.8);
  \draw (UL) -- ++( 120:1.0);
  \draw (UR) -- ++(  30:1.0);
  \draw[dotted] (UR)+( 30:0.8) to [bend left=30] ++( 60:0.8);
  \draw (UR) -- ++(  60:1.0);
  \draw (LR) -- ++( -30:1.0);
  \draw[dotted] (LR)+( -35:0.8) to [bend left=30] ++( -55:0.8);
  \draw (LR) -- ++( -60:1.0);
  \draw (LL) -- ++(-150:1.0);
  \draw[dotted] (LL)+( -150:0.8) to [bend left=30] ++( -120:0.8);
  \draw (LL) -- ++(-120:1.0);
\end{tikzpicture}
}
\end{align}
where the picture represents the reduced box integral, and $\Delta_{a b c d}$ is given in \eqref{Delta4m}. The fact that this factor cancels in the product of the box coefficient given by the quadruple cut \eqref{cuttoR} and the scalar integral is the main reason why we find convenient to use this basis. The factors on the left-hand side appear in front of any one-loop diagram and can be reabsorbed in the definition of the coupling. For completeness we remind the reader that
\begin{align}
 r_{\Gamma}&= \frac{\Gamma^2(1-\eps)\Gamma(1+\eps)}{\Gamma(1-2\eps)}  \ .
\end{align}
We also list the expression of the reduced box integrals that are needed for our computations:
\begin{align}
\raisebox{-.47\height}{
 \begin{tikzpicture}[thick,scale=0.65]
\drawULempty
\drawURempty
\drawLRempty
\drawLLempty
  \drawregionvariables{$x_c$}{$x_d$}{$x_a$}{$x_b$}{}
\draw (UL) -- (UR) -- (LR) -- (LL) -- (UL);
\draw (UL) -- ++( 135:1.0) node[above left=-2pt] {};
\draw (UR) -- ++(  45:1.0) node[above right=-2pt] {};
  \draw (LR) -- ++( -30:1.0);
  \draw[dotted] (LR)+( -35:0.8) to [bend left=30] ++( -55:0.8);
  \draw (LR) -- ++( -60:1.0);
\draw (LL) -- ++(-135:1.0) node[below left=-2pt] {};
\end{tikzpicture}
} = &\; -\frac{1}{\eps^2} \left((-x_{ac}^2)^{-\eps}+(-x_{bd}^2)^{-\eps}-(-x_{bc}^2)^{-\eps}\right) \\
 &\; + \Li\left(1-\frac{x_{bc}^2}{x_{ac}^2}\right)+\Li\left(1-\frac{x_{bc}^2}{x_{bd}^2}\right)+\frac12 \log^2\left(\frac{x_{ac}^2}{x_{bd}^2}\right)+\frac{\pi^2}{6} \nonumber \\
\raisebox{-.47\height}{
 \begin{tikzpicture}[thick,scale=0.65]
\drawULempty
\drawURempty
\drawLRempty
\drawLLempty
  \drawregionvariables{$x_c$}{$x_d$}{$x_a$}{$x_b$}{}
\draw (UL) -- (UR) -- (LR) -- (LL) -- (UL);
\draw (UL) -- ++( 150:1.0);
  \draw[dotted] (UL)+( 145:0.8) to [bend left=30] ++( 125:0.8);
  \draw (UL) -- ++( 120:1.0);
\draw (UR) -- ++(  45:1.0) node[above right=-2pt] {};
  \draw (LR) -- ++( -30:1.0);
  \draw[dotted] (LR)+( -35:0.8) to [bend left=30] ++( -55:0.8);
  \draw (LR) -- ++( -60:1.0);
\draw (LL) -- ++(-135:1.0) node[below left=-2pt] {};
\end{tikzpicture}\label{2measy}
} = &\; -\frac{1}{\eps^2} \left((-x_{ac}^2)^{-\eps}+(-x_{bd}^2)^{-\eps}-(-x_{bc}^2)^{-\eps}-(-x_{ad}^2)^{-\eps}\right) \\
 &\; + \Li\left(1-\frac{x_{ad}^2}{x_{ac}^2}\right) + \Li\left(1-\frac{x_{ad}^2}{x_{db}^2}\right) + \Li\left(1-\frac{x_{bc}^2}{x_{ac}^2}\right) + \Li\left(1-\frac{x_{bc}^2}{x_{db}^2}\right) \nonumber \\
 &\; - \Li\left(1-\frac{x_{ad}^2 x_{bc}^2}{x_{ac}^2 x_{db}^2}\right) + \frac{1}{2}\log^2\left(\frac{x_{ac}^2}{x_{db}^2}\right) \;, \nonumber \\
\raisebox{-.47\height}{
 \begin{tikzpicture}[thick,scale=0.65]
\drawULempty
\drawURempty
\drawLRempty
\drawLLempty
  \drawregionvariables{$x_c$}{$x_d$}{$x_a$}{$x_b$}{}
\draw (UL) -- (UR) -- (LR) -- (LL) -- (UL);
\draw (UL) -- ++( 135:1.0) node[above left=-2pt] {};
\draw (UR) -- ++(  30:1.0);
  \draw[dotted] (UR)+( 30:0.8) to [bend left=30] ++( 60:0.8);
  \draw (UR) -- ++(  60:1.0);
  \draw (LR) -- ++( -30:1.0);
  \draw[dotted] (LR)+( -35:0.8) to [bend left=30] ++( -55:0.8);
  \draw (LR) -- ++( -60:1.0);
\draw (LL) -- ++(-135:1.0) node[below left=-2pt] {};
\end{tikzpicture}
} = &\; -\frac{1}{\eps^2} \left(\frac12(-x_{ac}^2)^{-\eps}+(-x_{bd}^2)^{-\eps}-\frac12(-x_{bc}^2)^{-\eps}-\frac12(-x_{ab}^2)^{-\eps}\right) \\
 &\; + \Li\left(1-\frac{x_{bc}^2}{x_{bd}^2}\right) + \Li\left(1-\frac{x_{ab}^2}{x_{db}^2}\right) +\frac12 \log^2 \left(\frac{x_{ac}^2}{x_{bd}^2}\right) \nonumber \\
 &\; -\frac12 \log \left(\frac{x_{ac}^2}{x_{bc}^2}\right) \log \left(\frac{x_{ac}^2}{x_{ab}^2}\right) \nonumber 
\end{align}
In the main text we also use a F inside the diagram to indicate that we consider only the finite part of the one-loop integrals. By finite part we mean the previous expressions where the first line has been removed.

For triangles, we use a notation that is analogous to the box case
\begin{align}\label{reducedtriangles}
\frac{1}{2\pi^{2-\eps}r_{\Gamma}}\int \mathrm{d}^{4-2\epsilon} x_0 \; \frac{1}{x_{0a}^2 x_{0b}^2 x_{0c}^2 }= \frac{1}{\mathrm i \Delta_{a b c}} \raisebox{-.47\height}{
\begin{tikzpicture}[thick, scale=0.65]
\coordinate (T1) at ( 0.5 ,  0.87);
\coordinate (T2) at ( 0.5 , -0.87);
\coordinate (T3) at (-1.0 ,  0.0 );
\draw (T1) -- (T2) -- (T3) -- (T1);
\node[] (Tlbl3) at (-0.6 ,  0.8) {$x_a$};
\node[] (Tlbl1) at ( 1.2 ,  0.0) {$x_b$};
\node[] (Tlbl2) at (-0.6 , -0.8) {$x_c$};
\node[] (Tlbl0) at ( 0.0 ,  0.0) {};
\draw (T1) -- ++(  30:1.0);
  \draw[dotted] (T1)+( 30:0.8) to [bend right=30] ++( 60:0.8);
  \draw (T1) -- ++(  60:1.0);
\draw (T2) -- ++(  -30:1.0);
  \draw[dotted] (T2)+( -30:0.8) to [bend left=30] ++( -60:0.8);
  \draw (T2) -- ++(  -60:1.0);
\draw (T3) -- ++(  165:1.0);
  \draw[dotted] (T3)+( 165:0.8) to [bend right=30] ++( 195:0.8);
  \draw (T3) -- ++(  195:1.0);
\end{tikzpicture}
}
\end{align}
with
\begin{align}\label{Delta3m}
 \Delta_{a b c}=\sqrt{(x^2_{ac}-x^2_{bc}+x^2_{ab})^2-4x^2_{ab}x^2_{ac}}\ .
\end{align}
Notice that, for $x^2_{ab}=0$, this factor reads
\begin{align}\label{Deltatri2m}
 \Delta_{a\; a+1\; c}=x^2_{ac}-x^2_{a+1,c}\ .
\end{align}
The three possible cases are given by
 \begin{align}
 \raisebox{-.46\height}{
\begin{tikzpicture}[thick, scale=0.65]
\coordinate (T1) at ( 0.5 ,  0.87);
\coordinate (T2) at ( 0.5 , -0.87);
\coordinate (T3) at (-1.0 ,  0.0 );
\draw (T1) -- (T2) -- (T3) -- (T1);
\node[] (Tlbl3) at (-0.6 ,  0.8) {$x_a$};
\node[] (Tlbl1) at ( 1.2 ,  0.0) {$x_b$};
\node[] (Tlbl2) at (-0.6 , -0.8) {$x_c$};
\node[] (Tlbl0) at ( 0.0 ,  0.0) {};
\draw (T1) -- ++( 45:0.8);
\draw (T2) -- ++( -45:0.8);
\draw (T3) -- ++(  165:1.0);
  \draw[dotted] (T3)+( 165:0.8) to [bend right=30] ++( 195:0.8);
  \draw (T3) -- ++(  195:1.0);
\end{tikzpicture}
}&=\frac{(-x^2_{ac})^{-\eps}}{2\eps^2} \\
 \raisebox{-.5\height}{
\begin{tikzpicture}[thick, scale=0.65]
\coordinate (T1) at ( 0.5 ,  0.87);
\coordinate (T2) at ( 0.5 , -0.87);
\coordinate (T3) at (-1.0 ,  0.0 );
\draw (T1) -- (T2) -- (T3) -- (T1);
\node[] (Tlbl3) at (-0.6 ,  0.8) {$x_a$};
\node[] (Tlbl1) at ( 1.2 ,  0.0) {$x_b$};
\node[] (Tlbl2) at (-0.6 , -0.8) {$x_c$};
\node[] (Tlbl0) at ( 0.0 ,  0.0) {};
\draw (T1) -- ++( 45:0.8);
\draw (T2) -- ++(  -30:1.0);
  \draw[dotted] (T2)+( -30:0.8) to [bend left=30] ++( -60:0.8);
  \draw (T2) -- ++(  -60:1.0);
\draw (T3) -- ++(  165:1.0);
  \draw[dotted] (T3)+( 165:0.8) to [bend right=30] ++( 195:0.8);
  \draw (T3) -- ++(  195:1.0);
\end{tikzpicture}
}&=\frac{(-x^2_{bc})^{-\eps}-(-x^2_{ac})^{-\eps}}{2\eps^2} \label{2mtriangle} \\
\raisebox{-.46\height}{
\begin{tikzpicture}[thick, scale=0.65]
\coordinate (T1) at ( 0.5 ,  0.87);
\coordinate (T2) at ( 0.5 , -0.87);
\coordinate (T3) at (-1.0 ,  0.0 );
\draw (T1) -- (T2) -- (T3) -- (T1);
\node[] (Tlbl3) at (-0.6 ,  0.8) {$x_a$};
\node[] (Tlbl1) at ( 1.2 ,  0.0) {$x_b$};
\node[] (Tlbl2) at (-0.6 , -0.8) {$x_c$};
\node[] (Tlbl0) at ( 0.0 ,  0.0) {};
\draw (T1) -- ++(  30:1.0);
  \draw[dotted] (T1)+( 30:0.8) to [bend right=30] ++( 60:0.8);
  \draw (T1) -- ++(  60:1.0);
\draw (T2) -- ++(  -30:1.0);
  \draw[dotted] (T2)+( -30:0.8) to [bend left=30] ++( -60:0.8);
  \draw (T2) -- ++(  -60:1.0);
\draw (T3) -- ++(  165:1.0);
  \draw[dotted] (T3)+( 165:0.8) to [bend right=30] ++( 195:0.8);
  \draw (T3) -- ++(  195:1.0);
\end{tikzpicture}
}&=\Li(z)-\Li(\bar z)+\tfrac{1}{2}\log(z \bar z)\log\left(\frac{1-z}{1-\bar z}\right)
\end{align}
where, for the last integral, we used the variables \eqref{uandv}. One may be worried that the two-mass triangle is odd under the exchange of the two massive corners. In fact, this sign is compensated by the $\Delta$ factor \eqref{Deltatri2m}. Since we are expanding in terms of reduced integrals, we need to choose a convention and fix the sign of the coefficient accordingly. Using the convention \eqref{2mtriangle}, one can check that the coefficient \eqref{c2m}, which we determined by IR consistency, has the right sign to cancel the unwanted three-particle invariants in the IR divergent part of the form factor.

\section{Solution of the triple cut constraints}\label{solutiontriangle}
In this appendix we review some results of \cite{Forde:2007mi}, adapting them to our notation. In the conventions of section \ref{3mtriangle} we set $x_{bc}=K_1$ and $x_{ac}=K_2=q$. We define also the two massless projections
\begin{align}
 K_1^{\flat,\mu}&=\frac{K_1^{\mu}-\frac{K_1^2}{\gamma_{\pm}}K_2^{\mu}}{1-\frac{K_1^2 K_2^2}{\gamma_{\pm}^2}}\, ,  & K_2^{\flat,\mu}&=\frac{K_2^{\mu}-\frac{K_2^2}{\gamma_{\pm}}K_1^{\mu}}{1-\frac{K_1^2 K_2^2}{\gamma_{\pm}^2}}\, , 
\end{align}
where, using the variables \eqref{uandv},
\begin{align}
 \gamma_{+}&=q^2 (1-\bar z) \, , & \gamma_{-}&=q^2 (1- z)\, .
\end{align}
The two different values are associated to the two solutions of the kinematics constraints. In general the mapping between the two solutions is achieved by $z\leftrightarrow \bar z$. Consequently,
\begin{align}
 \frac{K_1^2}{\gamma_+}&=(1-z) \, , &  \frac{K_1^2}{\gamma_-}&=(1-\bar z) \, , & \frac{K_2^2}{\gamma_+}&=\frac{1}{1-\bar z} \, , & \frac{K_2^2}{\gamma_-}&=\frac{1}{1- z}\, .
\end{align}
We can now express the loop momenta in terms of these massless projections and their associated spinor variables $\lambda^{\alpha}_{K_i^{\flat}}$ and $\tilde\lambda^{\dot\alpha}_{K_i^{\flat}}$. Explicitly
\begin{align}
 \lambda^{\alpha}_{\ell_i}&=t \lambda^{\alpha}_{K_1^{\flat}}+\alpha_{i1} \lambda^{\alpha}_{K_2^{\flat}}\, , \\
 \tilde \lambda^{\dot \alpha}_{\ell_i}&=\frac{\alpha_{i2}}{t} \tilde\lambda^{\dot \alpha}_{K_1^{\flat}}+ \tilde \lambda^{\dot \alpha}_{K_2^{\flat}}\, ,
\end{align}
with coefficients 
\begin{align}
\alpha_{11}^+&=\frac{z (\bar z-1)}{z-\bar z}\, ,  & \alpha_{12}^+&= \frac{\bar z (z-1)}{(z-\bar z)(\bar z-1)}\, , \\
\alpha_{21}^+&=\frac{z (z-1)}{z-\bar z}\, ,  & \alpha_{22}^+&=\frac{\bar z}{z- \bar z}\, , \\
 \alpha_{31}^+&=\frac{\bar z (z-1)}{z-\bar z} \, , & \alpha_{32}^+&=\frac{z}{z-\bar z}\, .
\end{align}
The coefficients associated to the other solution can be found by exchanging $z\leftrightarrow \bar z$.

Notice that in the limit $t\to \infty$ all the $\lambda_{\ell_i}$ go to $\lambda_{K_1^{\flat}}$. Since the limit $t\to \infty$ is the one leading to the direct extraction of the three-mass triangle coefficient, the final result depends only on $K_1^{\flat}$. In particular, in  \eqref{Kflat} we used a rescaled version of it
\begin{align}\label{KflatK1flat}
 K^{\flat}=K_1^{\flat} \left(1-\frac{1-z}{1-\bar z}\right)\ .
\end{align}
The two are not equal, but all our results depend only on $\lambda_{K_1^{\flat}}$ and we can use the rescaling freedom to replace $\lambda_{K_1^{\flat}}\to \lambda_{K^{\flat}} $.

Nevertheless, one should be careful because  \eqref{c3mdci} depends also on the contractions $\braket{\ell_i \ell_j}$ and the subleading order as $t\to \infty$ becomes relevant in that case,
\begin{align}
 \braket{\ell_1 \, \ell_2}_+&= t z \braket{K_1^{\flat} \, K_2^{\flat}} \, , \\
 \braket{\ell_1 \, \ell_3}_+&= t \braket{K_1^{\flat} \, K_2^{\flat}} \, , \\
 \braket{\ell_2 \, \ell_3}_+&= t (1-z)\braket{K_1^{\flat} \, K_2^{\flat}} \, .
\end{align}
Once more, the other solution is obtained with the replacement $z\to \bar z$. Using these expressions it is  easy to go from \eqref{c3mdci} to \eqref{c3msol}.
In our alternative expression for the coefficient, \eqref{c3mintermediate}, as well as \eqref{Rrs}, depends on the loop momenta only through $\lambda_{\ell_2}$ and this allows to use straightforwardly the replacement \eqref{l2Kflat}.

\section{Some dual conformal variations}\label{variations}
Here we consider explicit variations under dual conformal transformations of the function $g(u,v)$ defined in \eqref{gdef}. We start from \eqref{uvvar} and we derive
\begin{align}
\mathsf{K}^\mu z &= \frac{2(z-1)z}{z-\bar{z}} \left((1-\bar{z})x_{ab}^{\mu} - \bar{z}x_{bc}^{\mu}\right) \;, \cr
\mathsf{K}^\mu \bar{z} &= \frac{2(1-\bar{z})\bar{z}}{z-\bar{z}} \left((1-z)x_{ab}^{\mu} - zx_{bc}^{\mu}\right) \;.
\end{align}
The variation of $\D=|z-\bar z|$ follows immediately
\begin{align}
 \mathsf{K}^\mu \D= \frac{2[v(1+u-v)x_{ab}^{\mu}-u(1-u+v) x_{bc}^{\mu}]}{\D}\ , 
\end{align}
and it is clearly antisymmetric under the exchange \eqref{exchange}. Also the variation of $F^{3\mathrm{m}}$ in \eqref{F3m} is easily computed
\begin{align}
\mathsf{K}^\mu F^{3\mathrm{m}}(z,\bar z) = -\frac{\log u}{\Delta} \left((u+v-1)\,x_{ab}^\mu + 2u\,x_{bc}^\mu\right) + \frac{\log v}{\Delta} \left((u+v-1)\,x_{bc}^\mu + 2v\,x_{ab}^\mu\right) \;, 
\end{align}
and is antisymmetric as expected. The last ingredient in $g(u,v)$ is $\sqrt{uv}$,  whose variation is simply
\begin{align}
 \mathsf{K}^\mu \sqrt{uv} = (x_{ab}^{\mu}-x_{bc}^{\mu}) \sqrt{u v} \ . 
\end{align}
Therefore we have shown with an explicit  computation  that the variation of $g(u,v)$ under dual special conformal transformations is antisymmetric under the exchange \eqref{exchange}.

\pagebreak
	\bibliographystyle{utphys}
	\bibliography{remainder}

\end{document}